\documentclass[12pt]{article}
\usepackage{epsf}
\usepackage{aas_macros}
\usepackage{amssymb}
\usepackage{amsmath}
\usepackage{braket}
\usepackage{mathrsfs}
\usepackage{mathtools}
\usepackage{array}
\usepackage{fancyhdr}
\usepackage[dvipdfmx]{graphicx}
\usepackage{color}
\usepackage{cite}
\usepackage{bm}
\usepackage{url}
\usepackage{fancyhdr}
\usepackage[colorlinks,citecolor=blue]{hyperref}
\renewcommand{\thefootnote}{\#\arabic{footnote}}

\renewcommand{\thefootnote}{\fnsymbol{footnote}}
\def\thefootnote{\fnsymbol{footnote}}

\makeatletter

\@addtoreset{equation}{section}
\makeatother

\newcommand{\al}{\alpha}

\newcommand{\de}{\delta}
\newcommand{\De}{\Delta}

\newcommand{\PBH}{\mathrm{PBH}}

\newcommand{\x}{{\bm x}}
\newcommand{\y}{\bm y}

\newcommand{\D}{d}

\begin{document}

\begin{titlepage}

\begin{center}

\vskip .75in

{\Large \bf Angular correlation  as a novel probe \vspace{2mm} \\ of supermassive primordial black holes }

\vskip .75in

{\large
Takumi Shinohara$\,^1$,  Teruaki Suyama$\,^2$ and  Tomo~Takahashi$\,^3$
}

\vskip 0.25in

{\em
$^{1}$Graduate School of Science and Engineering, Saga University, Saga 840-8502, Japan
\vspace{2mm} \\
$^{2}$Department of Physics, Tokyo Institute of Technology, 2-12-1 Ookayama, Meguro-ku,
Tokyo 152-8551, Japan
\vspace{2mm} \\
$^{3}$Department of Physics, Saga University, Saga 840-8502, Japan
}

\end{center}
\vskip .5in

\begin{abstract}

We investigate the clustering property of primordial black holes (PBHs) in a scenario where PBHs can explain the existence of supermassive black holes (SMBHs) at high redshifts. We analyze the angular correlation function of PBHs originating from fluctuations of a spectator field which can be regarded as a representative model to explain SMBHs without conflicting with the constraint from the spectral distortion of cosmic microwave background.  We argue that the clustering property of PBHs can give a critical test for models with PBHs as the origin of SMBHs and indeed  show that the spatial distribution of PBHs in such a scenario is highly clustered, which suggests that those models may be disfavored from observations of SMBHs although a careful comparison with observational data would be necessary.

\end{abstract}

\end{titlepage}

\renewcommand{\thepage}{\arabic{page}}
\setcounter{page}{1}
\renewcommand{\thefootnote}{\#\arabic{footnote}}
\setcounter{footnote}{0}

\section{Introduction \label{sec:intro}}

Primordial black holes (PBHs), which are considered to be created in the early Universe \cite{Hawking:1971ei,Carr:1974nx,Carr:1975qj}, have been attracting attention, particularly since LIGO detected gravitational wave signals \cite{Abbott:2016blz}, which can be interpreted as the ones from the  merger of PBH binary \cite{Bird:2016dcv,Clesse:2016vqa,Sasaki:2016jop}.\footnote{
There are also several astrophysical explanations for the detected BH binaries. 
They include field formation \cite{Belczynski:2016obo},
chemically-homogeneous evolution \cite{Mandel:2015qlu}, 
and dynamical formation in dense stellar clusters \cite{Rodriguez:2016kxx}.
} PBHs have also been discussed as a possible dark matter (DM) candidate, which has also motivated the study of PBHs. Although the abundance of PBHs is severely constrained, they can constitute all DM in the Universe in some mass range, or some fraction of DM can be explained by PBHs \cite{Carr:2020gox}. 

Another motivation to investigate PBHs is the issue of supermassive black holes (SMBHs) observed at very high redshifts. 
The masses of SMBHs found in the redshift $z \simeq 5 - 7$ are $10^7 \, M_\odot - 10^{10} \, M_\odot$, which is considered to be challenging to be formed by some astrophysical processes (e.g., see \cite{Woods:2018lty} and reference therein). 
Actually, assuming that the BH continues the Eddington-limited accretion, 
the time for the BH initially having mass $M_{\rm ini}$ 
to reach the mass $M_{\rm fin}$
is given by \cite{Woods:2018lty}
\begin{equation}
    t_{\rm grow} \simeq 5\times 10^{-2} \left( \frac{\epsilon_r}{0.1} \right)
    \ln \left( \frac{M_{\rm fin}}{M_{\rm ini}} \right) ~{\rm Gyr},
\end{equation}
where $\epsilon_r$ (for thin disk accretion $\sim 0.1$ \cite{Shakura:1972te}) is the radiative efficiency.
For $M_{\rm fin} \gg M_{\rm ini}$ (for instance, Pop III stars would have $M_{\rm ini}\simeq 10 M_\odot - 10^3 M_\odot$ \cite{Hirano:2013lba}), 
$t_{\rm grow}$ is barely less than the age of the Universe at $z=5 - 7$.
It is not yet clear if such an efficient accretion can persistently operate to
make stellar-mass BHs grow rapidly into SMBHs 
by the redshift $z\sim 7$ (e.g., \cite{Johnson:2006gd,2018MNRAS.480.3762S}). 
Another possibility is the direct collapse scenario (e.g., \cite{Bromm:2002hb, Hosokawa:2013mba})
in which very massive stars
$\sim 10^5 M_\odot$ were formed directly from the gravitational collapse 
of the gas cloud. 
It is worth noting that the quasars at high redshifts can be also explained as the hypermassive starburst clusters \cite{Kroupa:2020sru}.  In this paper, we assume that the observed quasars at high redshifts are SMBHs.

In addition to the astrophysical scenarios explaining such SMBHs, 
PBHs as progenitor of SMBHs are an alternative possibility and have been studied in the literature \cite{Duechting:2004dk,Kawasaki:2012kn,Nakama:2016kfq,Hasegawa:2017jtk,Kawasaki:2019iis,Kitajima:2020kig}. 
While the production of PBHs needs models beyond the standard cosmological 
scenario (e.g., see \cite{Allahverdi:2020bys}), one of the attractive features 
of the PBH scenario is that there is a simple physical mechanism to produce PBHs 
having initial masses in the mass range of SMBHs or seeds of SMBHs; namely,
preparing the primordial curvature perturbations with the ${\cal O}(1)$ amplitude 
at the corresponding length scale inevitably leads to the formation of PBHs, for which 
initial PBH masses are determined by the (comoving) wave number of the primordial 
perturbations at which the power spectrum is enhanced.

In this paper, 
we consider PBHs in the mass range of $10^4 M_\odot - 10^{10} M_\odot$.
As the lower limit of the initial PBH masses, we take $10^4~M_\odot$
since Population III remnants are expected to form BHs up to $\sim 10^3 M_\odot$ \cite{Haemmerle:2020iqg}.
Replacing such seed BHs originating from the Pop III stars by PBHs
of the same initial masses 
does not sound appealing although it is logically possible.
For PBHs with the initial masses of $\sim 10^4 M_\odot$,
although it is not clear if the astrophysical processes such as accretion 
made those PBHs grow rapidly and achieved the masses of the SMBHs observed at high redshifts,
we assume that such processes operated successfully.
On the other hand, if PBHs were in the SMBH-mass range initially, 
a very efficient accretion  
is not needed to explain the SMBHs. 
In any case, as we will see later, our main result that the PBH models we consider
predict very high degrees of the clustering of PBHs over wide length scales holds true for any mass range of our interest.
In this sense, whether PBHs started from around $10^4 M_\odot$ or had masses in the SMBH mass range
from the outset is not crucial for our conclusion.

However, it is known that if PBHs with masses $10^4 M_\odot - 10^{13} M_\odot$ are formed 
from the Gaussian primordial curvature perturbation to explain SMBHs, 
such primordial perturbations inevitably produce the spectral distortion of the 
cosmic microwave background (CMB) which largely exceeds the upper limit obtained by the COBE/FIRAS experiment \cite{Kohri:2014lza}. 
Therefore, one needs to consider a scenario where PBHs are created from highly non-Gaussian perturbations \cite{Kohri:2014lza,Nakama:2016kfq}, which are realized in some concrete models \cite{Nakama:2016kfq,Hasegawa:2017jtk,Kawasaki:2019iis,Kitajima:2020kig}. 
Importantly, as it will be explained in more detail in the subsequent sections,
when PBHs are produced from such non-Gaussian perturbations, clustering of PBHs inevitably occurs. 
Hence the clustering property of PBHs can be a crucial test of a scenario where PBHs can explain SMBHs at high redshifts. 

In this paper, we discuss the formalism to calculate two-point angular 
correlation function of PBHs to study its clustering properties. 
Then we apply it to investigate the clustering of PBHs in models where highly non-Gaussian fluctuations from a spectator field produce PBHs as a seed of SMBHs.  
We stress that our formalism can be applied to a broad class of models where PBHs are produced from highly non-Gaussian perturbation, and argue that the PBH angular correlation function gives a critical test to a scenario of PBHs as the origin of SMBHs. 

The structure of this paper is as follows. 
In the next section, we will explain a common feature present in the proposed models,
which provides a motivation to theoretically derive the PBH angular correlation function.
After this, we explicitly calculate the PBH angular correlation function 
in a scenario where PBHs as a seed of SMBHs are created from highly non-Gaussian fluctuations sourced by a spectator field. 
We also discuss how the clustering properties depend on the details of the model. 
Then in the final section, we conclude our paper.

\section{PBHs from a  spectator field \label{sec:model}}
In this section, we describe one important feature common to the 
proposed models in the literature \cite{Nakama:2016kfq,Hasegawa:2017jtk,Kawasaki:2019iis,Kitajima:2020kig}
explaining SMBHs by PBHs and provide the motivation to analyze 
the angular correlation of such PBHs.
As mentioned in the Introduction, one of the advantages of the PBH scenario is that
PBHs can have masses in the SMBH-mass range from the outset and
we do not need to resort to the efficient accretion.
Yet, PBHs may grow to some extent by accretion and it is possible that initial PBHs are
lighter than the SMBHs.
The uncertainty about how much the PBHs can increase their mass during the cosmological history propagates into the uncertainty of the (comoving) wave number 
of the primordial curvature perturbations which collapse to the PBHs.
As we will demonstrate in Sec.~\ref{sec:results},
the PBH correlation function depends on the wave number ($k_{\rm max}$ in our notation),
but our main conclusion
that the PBH angular correlation function is very large on small angular scales
is universal for any values of $k_{\rm max}$.

Irrespective of whether the initial PBH masses are about $10^4 M_\odot$ or much closer to the masses of the observed SMBHs,
if we only aim at explaining the observed abundance of SMBHs,
this can be easily achieved simply by
having the primordial curvature perturbations with ${\cal O}(1)$ amplitude at the required length scales.
What is rather non-trivial is that we need to achieve this without contradicting with the non-observations of the CMB spectral distortions.
Actually, the variance of the primordial curvature perturbations is known to become much bigger than 
the upper limit derived from the non-observations of the CMB spectral distortions if such PBHs are produced from (almost) the Gaussian curvature perturbation \cite{Kohri:2014lza}.
Therefore, if SMBHs originate from PBHs which were formed by the gravitational collapse
of the primordial curvature perturbations, the primordial curvature perturbations must be highly non-Gaussian
such that the variance is small enough to evade the CMB constraints while, at the same time,
the probability to realize the ${\cal O}(1)$ amplitude of the perturbation to form PBHs 
is large enough to explain the
observed abundance of SMBHs (the comoving number density of the SMBHs is known to be about $ 1 {\rm Gpc}^{-3}$ \cite{Fan:2003wd}).

\begin{figure}[t]
  \begin{center}
    \includegraphics[clip,width=15.0cm]{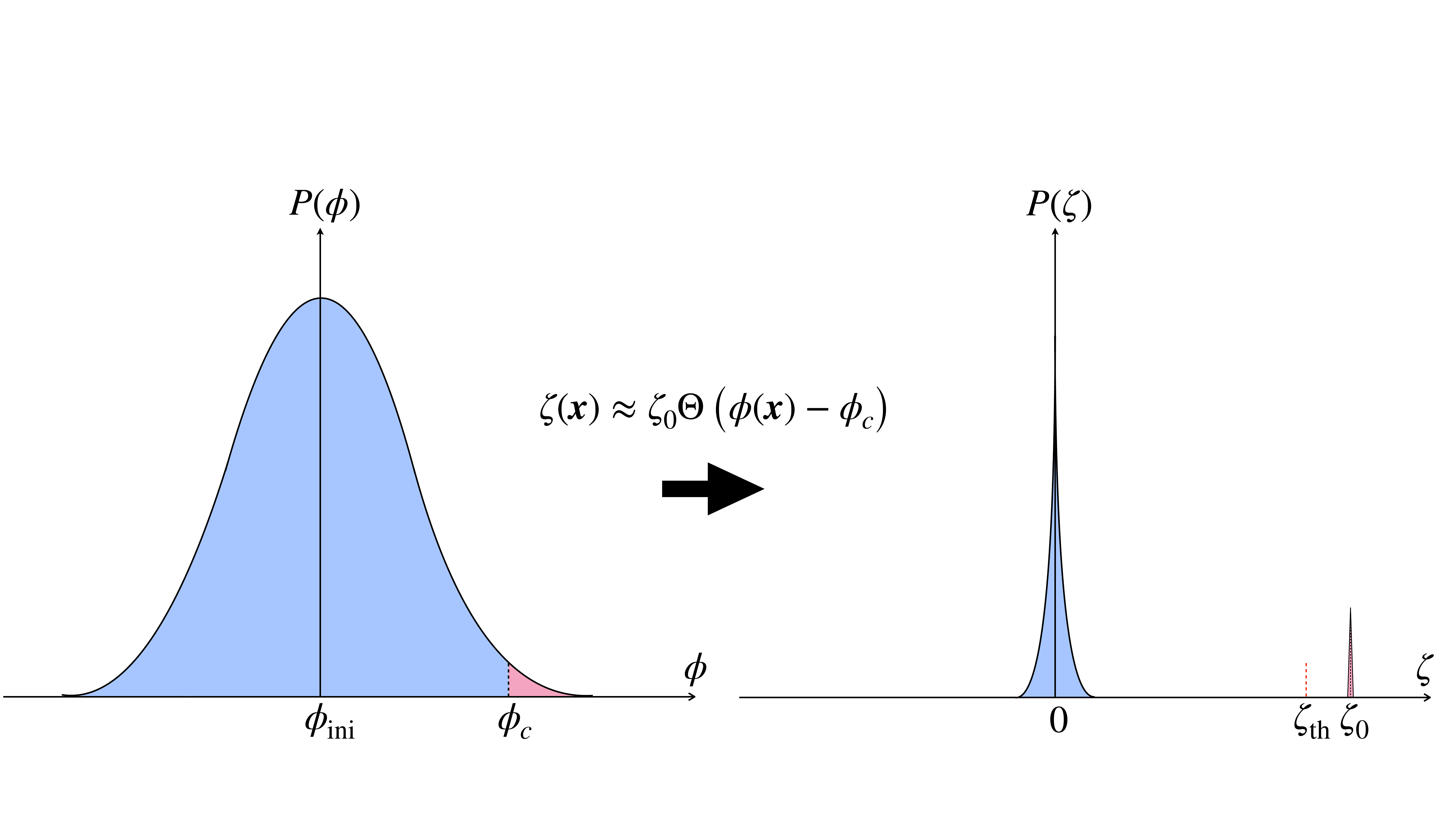}
    \caption{Left panel: (Gaussian) probability distribution  of the spectator field $\phi$
    as a result of the quantum fluctuations generated during inflation.
    Right panel: schematic picture of probability distribution of the primordial curvature perturbation $\zeta$.
    Hubble patches having $\phi <\phi_c$ produce little of the curvature perturbation and only
    the patches having $\phi >\phi_c$ acquire the amplitude $\zeta_0$ greater than the threshold $\zeta_{\rm th}$ 
    to collapse into PBHs.}
    \label{pdf}
  \end{center}
\end{figure}

In \cite{Nakama:2016kfq,Hasegawa:2017jtk,Kawasaki:2019iis,Kitajima:2020kig}, 
inflationary models have been proposed to realize
such highly non-Gaussian perturbations.
Although the concrete scenarios are different from each other,  
what is common among them is that a light scalar field $\phi$ (spectator field) undergoes stochastic motion during inflation and only the patches where the field value evaluated at the end of inflation
exceeds a threshold $\phi_c$ acquire ${\cal O}(1)$ curvature perturbations (by the end of inflation
or afterwards depending on the underlying models)
and the curvature perturbations in the other patches are negligibly small.\footnote{
Alternative scenarios include nucleation of bubbles by quantum tunneling during inflation \cite{Deng:2018cxb,Deng:2020pxo,Deng:2020mds} and transient constant-roll inflation \cite{Atal:2019cdz,Atal:2020yic}.
We do not consider this class of models in our paper.
}
Denoting by $\phi ({\bm x})=\phi (t_{\rm end},{\bm x})$ the value of the scalar field at the end of inflation 
smoothed over the scales relevant to PBHs,
the curvature perturbation produced in this class of models can be approximately written as
\begin{equation}
\label{approx-zeta}
\zeta ({\bm x}) \approx \zeta_0 \Theta (\phi ({\bm x})-\phi_c).
\end{equation}
Here $\Theta$ is the step function,
$\zeta_0$, which should be larger than the threshold $\zeta_{\rm th}$ for the PBH formation \cite{Shibata:1999zs}, 
is the amplitude of the curvature perturbation in the patches that collapse to PBHs at the time of the horizon reentry, 
and $\phi_c$ is the critical value whose explicit value depends on the underlying models.\footnote{
Precisely speaking, in the models studied in \cite{Hasegawa:2017jtk,Kawasaki:2019iis,Kitajima:2020kig}, 
it is the absolute value of $\phi$ that enters in Eq.~(\ref{approx-zeta}). 
When the PBH formation is very rare and the dynamics of $\phi$ is symmetric for $\phi \to -\phi$, which is actually the case, patches having $\phi>\phi_c$
will not be correlated with the ones having $\phi< -\phi_c$ and 
we can virtually use Eq.~(\ref{approx-zeta}) to derive the correlation function
of $\zeta$.
}
As it is evident from this equation,
the distribution of $\zeta$ is highly non-Gaussian 
and the non-observations of the CMB spectral distortion can be easily met due to the rareness of the patches having $\phi > \phi_c$ (see Fig.~\ref{pdf}). 

In this way, we can construct consistent models of PBHs which can explain SMBHs
without conflicting with the observational upper limit on the CMB spectral distortion. 
Therefore, it is desirable to consider additional observables which can be used to test these models.
To this end, let us recall that the approximate equation of motion of $\phi$
at each Hubble patch during inflation is given by (e.g., \cite{Starobinsky:1994bd})
\begin{equation}
\label{langevin}
{\dot \phi} (t)=-\frac{m^2}{3H} \phi (t)+\xi (t),
\end{equation}
where $m$ is the mass of $\phi$, $\xi (t)$ is the Gaussian noise coming from the transition of the sub-Hubble quantum 
fluctuations to the super-Hubble ones
and its two-point function is given by $\langle \xi (t) \xi (t') \rangle=\frac{H^3}{4\pi^2} \delta (t-t')$.
Let us suppose that $\phi$ at the  time $(t=0)$ when our observable Universe exited the Hubble horizon was $\phi_{\rm ini}$.
In our observable Universe,
after this time, $\phi$ moves randomly according to Eq.~(\ref{langevin})
and the distribution of $\phi$ becomes Gaussian with its mean being $\phi_{\rm ini}$.
Roughly speaking, a PBH of mass $M_{\rm PBH}$ is formed (after inflation) in the Hubble patch if
$\phi (t)$ evaluated at the time when the length scale corresponding to $M_{\rm PBH}$ 
exited the Hubble horizon exceeds $\phi_c$.
The variance of $\phi$ at $t (>0)$ is given by
\begin{equation}
\langle {(\phi (t)-\phi_{\rm ini})}^2 \rangle =
\frac{3H^4}{8\pi^2 m^2} \left( 1-
e^{-\frac{2m^2}{3H}t} \right) \equiv \sigma_t^2.
\end{equation}
Thus, the fraction of such PBHs at their formation time is given by 
\begin{equation}
\beta \equiv
\int_{\phi_c}^\infty P(\phi) d\phi
\simeq \frac{1}{\sqrt{2\pi}}
\frac{\sigma_t}{\phi_c}
\exp \left( -\frac{\phi_c^2}{2 \sigma_t^2} \right),
\end{equation}
if the PBH formation is a rare process: $\sigma_t \ll \phi_c-\phi_{\rm ini}$.
Here $P(\phi)$ is the (Gaussian) probability distribution of $\phi$.
Since $\phi (t)$ is determined by the random motion accumulated until that time,
if a given Hubble patch has the field value larger than $\phi_c$, 
nearby patches, which experienced similar evolutionary history of $\phi$, 
would have more chance to have the field value larger than $\phi_c$ compared to the average.
This implies that patches having $\phi > \phi_c$ do not obey the Poisson distribution but are spatially clustered.
Consequently, PBHs produced in this way would be clustered as well, which motivates us
to study the correlation function of PBHs with an expectation that 
this provides a critical observable to test PBHs as the origin of the SMBHs.

Based on the above consideration, 
the purpose of this paper is to derive the two-point (angular) 
correlation function of PBHs originating from the curvature perturbations of the form given by Eq.~(\ref{approx-zeta})\footnote{
It should be noted that the two-point correlation function of the fluctuations exceeding the large threshold
has been derived for the general Gaussian fluctuations in \cite{Ali-Haimoud:2018dau}. At the formal level, we recover the result given in \cite{Ali-Haimoud:2018dau},
but we furthermore consider the result in the class of SMBH models of our interest and give the predictions for its angular correlation function.
}.
The above argument is suggestive but just qualitative, and in the next section we will provide a
more mathematically rigorous analysis.
Our crucial assumptions are that the mass of the spectator field is much smaller than the Hubble parameter during inflation, and self-interactions or interactions with any other fields can be neglected, both of which are satisfied in the models studied in \cite{Nakama:2016kfq,Hasegawa:2017jtk,Kawasaki:2019iis,Kitajima:2020kig}. 
Another important condition is that the curvature perturbation large enough to generate PBHs is produced only in the regions with $\phi (\bm x) > \phi_c$ [see Eq.~\eqref{approx-zeta}]. 
Apart from these general requirements, we do not need to specify the underlying inflation models.
Thus, our results presented in this paper can be applied to any other models  as long as the above assumptions are satisfied. We stress that our formalism can be utilized to investigate PBH clustering not only in SMBH mass but also in any other mass ranges.

Before closing this section, we mention that the PBHs from the curvature perturbations possessing a local-type non-Gaussianity 
of the form $\zeta=\zeta_{\rm g}+ \frac{3}{5} f_{\rm NL}\zeta_{\rm g}^2$ ($\zeta_{\rm g}$:~Gaussian field, $f_{\rm NL}$:~free parameter), 
which has been widely studied in the literature, have been known to cluster \cite{Tada:2015noa,Young:2015kda,Suyama:2019cst,Matsubara:2019qzv}.
On the other hand, the quantitative analysis of the clustering of the PBHs for the non-Gaussianity 
of the type (\ref{approx-zeta}) has not been performed in the literature.

\section{Formalism \label{sec:formalism}}

In this section, we give the formalism to study the clustering properties of PBHs as a possible explanation of SMBHs at high redshifts. For this purpose, we investigate the angular correlation function of PBHs.  
Since the angular correlation function can be calculated from the PBH two-point correlation function, 
we first briefly summarize the formalism to calculate the two-point correlation function of PBHs based on the functional integration approach \cite{Suyama:2019cst} (see also \cite{Matarrese:1986et,Franciolini:2018vbk}). 
Then we introduce the PBH angular correlation function as a measure to test the clustering properties of PBHs.

\subsection{PBH correlation function \label{sec:2pt_corr_func}}
  
Here we take the functional integration or the path integral approach to calculate 
the two-point correlation function of PBHs based on \cite{Suyama:2019cst}. 
We refer the readers to \cite{Suyama:2019cst} for details. 

As described in the previous section, we consider a scenario with a spectator field $\phi$ where a PBH is formed in the region satisfying the condition $\phi - \phi_{\rm ini} > \phi_c$. 
It should be understood that $\phi$ appearing in this condition is the one evaluated 
at the time of the end of inflation
and the fate of the given region, i.e., whether it later collapses to PBH or not, is determined by the
{\it initial condition} $\phi ({\bm x})=\phi ( t_{\rm end},{\bm x})$ evaluated at the inflation end. 
With this framework, the probability that a PBH is produced at the position $\bm x$ can be written as 
\begin{equation}
\label{eq:P_1}
{P}_1(\x) = \int [D\phi]\ P[\phi]\ \int^\infty_{\phi_c} d\alpha\ \delta_D (\phi(\x)-\alpha) \,,
\end{equation}
where $P [\phi]$ is the probability functional having a configuration $\phi ({\bm x})$, 
$\de_D$ is the Dirac's $\delta$-functions, and
\begin{equation}
\left[ D\phi \right]=\prod_{i=1} d \phi(\x_i). 
\end{equation}
In the same way, the probability that PBHs are created at the positions $\x_1$ and $\x_2$ is given by 
\begin{equation}
\label{eq:P_2}
{P}_2(\x_1,\x_2)=	\int [D\phi]\ P[\phi]\ \int^\infty_{\phi_c}\D\alpha_1\ \de_{D}(\phi(\x_1)-\alpha_1)\ \int^\infty_{\phi_c}\D\alpha_2\ \de_{D} (\phi(\x_2)-\alpha_2) \,.
\end{equation}

By using the integral form of the  $\delta$-function 
\begin{eqnarray}
\label{eq:delta_func}
\de_{D}(\phi-\alpha)=\int_{-\infty}^{\infty} \frac{\D q}{2\pi}\ e^{iq(\phi-\al)} \,,
\end{eqnarray}
one can express $P_1(\x)$ and $P_2 (\x_1, \x_2)$ as 
\begin{eqnarray}
\label{eq:P_1_2}
 &\ &{P}_1(\x)= \int [{D}\phi]\ P[\phi]\ \int^\infty_{\phi_c}\D\al\ \int^\infty_{-\infty}\frac{\D q}{2\pi}\ e^{-iq\al}\ e^{iq\phi(\x)} \,,\\  [8pt]
 \label{eq:P_2_2} 
 &\ &{P}_2(\x_1,\x_2)= \int [{D}\phi]\ P[\phi]\ \int^\infty_{\phi_c}\D\al_1\ \int^\infty_{\phi_c}\D\al_2\ \int^\infty_{-\infty}\frac{\D q_1dq_2}{(2\pi)^2}
e^{ -iq_1\al_1-iq_2\al_2} e^{iq_1\phi(\x_1)+iq_2\phi(\x_2)}\,. \notag \\ 
\end{eqnarray}

Now we define the generating functional for the general source $J(\x)$ as 
\begin{eqnarray}
 \label{eq:generating_func}
 Z[J]&\equiv& \int [{D}\phi]\ P[\phi]\ \exp\left[i\int \D^3y\ J(\y)\phi(\y) \right] = \left<\exp\left[i\int \D^3y\ J(\y)\phi(\y) \right] \right>.
 \end{eqnarray}
From this equation, it is known that the connected part of the $n$-point correlation function 
for $\phi$ can be written in terms of $Z [J]$ as
\begin{eqnarray}
\label{eq:connected_npoint_fun}
\xi^{(n)}_{\phi({c})}(\x_1,\cdots,\x_{n})
=
\left<\phi(\x_1)\cdots\phi(\x_2)\right>_{c} 
= 
\frac{1}{i}\ \left.\frac{\de^{n}\ln Z[J]}{\de J(\x_1)\cdots \de J(\x_{n})}\right|_{J=0} \,.
\end{eqnarray}
It is straightforward to invert this equation as
\begin{eqnarray}
\label{eq:logZ}
\ln Z[J]=\sum^\infty_{n=2}\frac{i^n}{n!}\int\left[\prod^{n}_{i=1}\D^3y_i \,  J(\y_i)\right]\ \xi^{(n)}_{\phi({c})}(\y_1,\cdots,\y_n) \,.
\end{eqnarray}

By the generating functional, we can formally express $P_1$ and $P_2$ in a compact manner.
By setting $J(\y)$ as 
\begin{eqnarray}
\label{eq:J_P1}
 J(\y)\equiv q \, \de_{D}(\x-\y) \,,
\end{eqnarray}
one can show that $ P_1(\x)$ can be given by
\begin{eqnarray}
\label{eq:P1_with_J}
{P}_1(\x) & =&\int^\infty_{\phi_{c}}\ \D\al\ \int^\infty_{-\infty}\frac{\D q}{2\pi}\ e^{-iq\al}\ Z[q\de_{D}(\x-\y)]  \,.
\end{eqnarray}
We can also calculate ${P}_2 (\x_1, \x_2)$ in a similar way. By setting 
\begin{eqnarray}
\label{eq:J_P2}
 J(\y)\equiv q_1 \, \de_{D}(\x_1 - \y) +  q_2 \, \de_{D}(\x_2-\y)  \,,
\end{eqnarray}
$P_2$ is written as
\begin{equation}
\label{Z-P2}
    {P}_2(\x_1,\x_2)=\int^\infty_{\phi_{c}}\ \D\al_1\
    \int^\infty_{\phi_{c}}\ \D\al_2 \int^\infty_{-\infty}\frac{\D q_1 \D q_2}{4\pi^2}\ e^{-iq_1\al_1-iq_2\al_2}\ Z[q_1 \de_{D}(\x_1-\y)+q_2 \de_{D}(\x_2-\y)]. 
\end{equation}

Here we assume that fluctuations of $\phi$ are Gaussian. 
In this case, the connected $n$-point correlation functions with $n \ge 3$ vanish 
and $\ln Z [J]$ becomes
\begin{eqnarray}
\label{eq:J_gaussian}
\ln Z[J] = -\frac12 \int d^3y_1  d^3y_2 J(\y_1) J(\y_2) \xi_\phi (\y_1, \y_2),
\end{eqnarray}
where we have dropped the subscript $(c)$ and the superscript $(2)$ for notational simplicity. 
By putting this into Eq.~\eqref{eq:P1_with_J}, and introducing the following two variables 
defined by
\begin{eqnarray}
\sigma^2  \equiv \xi_\phi (\x, \x), 
\qquad
\nu\equiv\frac{\phi_c}{\sigma} \,, 
\end{eqnarray}
one can show that the one-point probability of PBHs can be written as
\begin{eqnarray}
\label{eq:P_1_3}
{P}_1(\x)  
&=&\frac{1}{2}\ \mathrm{erfc}\!\left(\frac{\nu}{\sqrt{2}}\right) \simeq  \frac{e^{-\nu^2/2}}{\sqrt{2\pi}\ \nu} \,,
\end{eqnarray}
where erfc is the
complementary error function. 
Since we consider the formation of PBHs which is a rare process [see discussion around Eq.~(\ref{eq:beta})],  
we have adopted a high peak approximation where  $\nu  = \phi_c / \sigma \gg 1 $, 
which is used in the second equality in the above equation.

Plugging Eq.~(\ref{eq:J_gaussian}) into Eq.~(\ref{Z-P2}), we obtain
\begin{equation}
    P_2 ({\bm x}_1, {\bm x}_2)=
    \int_{\phi_c}^\infty d \alpha_1 \int_{\phi_c}^\infty d \alpha_2
    \int_{-\infty}^\infty \frac{dq_1 dq_2}{4\pi^2}
    e^{-iq_1 \alpha_1-i q_2 \alpha_2-\frac{1}{2} q_1^2
    \sigma^2-q_1 q_2 \xi_\phi ({\bm x}_1,{\bm x}_2)-\frac{1}{2} q_2^2 \sigma^2}.
\end{equation}
For notational simplicity, let us introduce a dimensionless quantity $\epsilon$ by
\begin{equation}
\label{def-epsilon}
 \epsilon =   \frac{\xi_\phi ({\bm x}_1,{\bm x}_2)}{\sigma^2}.
\end{equation}
By definition, $\epsilon \le 1$.
Integration over $q_1, q_2$ yields, after changing the integration variables, 
\begin{equation}
    P_2 ({\bm x}_1, {\bm x}_2)=
    \frac{\sqrt{1-\epsilon^2}}{2\pi}
    \int_{\frac{\nu}{\sqrt{1-\epsilon^2}}}^\infty dv_1
    \int_{\frac{\nu}{\sqrt{1-\epsilon^2}}}^\infty dv_2
    ~\exp \left( -\frac{1}{2} (v_1^2-2\epsilon v_1 v_2 +v_2^2 ) \right) \,.
\end{equation}
The integration over $v_2$ can be written in terms of the
complementary error function  as
\begin{equation}
\label{P2-2}
    P_2 ({\bm x}_1, {\bm x}_2)=
    \frac{1}{2\sqrt{2\pi}} \int_\nu^\infty dz
    ~e^{-\frac{1}{2}z^2} {\rm erfc} \left(
    \frac{\nu-\epsilon z}{\sqrt{2(1-\epsilon^2)}} \right).
\end{equation}
Up to this point, the expression for $P_2$ is mathematically exact.
Although we can, in principle, obtain the accurate correlation function 
by performing the integration numerically, 
it is possible to perform the integration analytically under some approximation,
which not only helps us capture the basic behaviors of the correlation function easily but also reduces the computation time to obtain the final angular correlation function which involves multiple integration. 
To this end, we notice that the variable $z$ is greater than $\nu$.
Thus, the argument of the ${\rm erfc}$ function becomes much bigger than unity
for $\nu \sqrt{1-\epsilon} \gg 1$.
In what follows, we assume $\nu\sqrt{1-\epsilon} \gg 1$ is satisfied.
This condition is violated at the short distance ${\bm x}_1 \to {\bm x}_2$ for which $\epsilon \to 1$.
Such a short distance is not observationally relevant.
Furthermore, dynamical interactions among PBHs may become important at such short distances,
by which the correlation function that can be  measured may differ from the primordial one we compute  
(see Sec.~\ref{app-fraction}).
Thus, the condition $\nu \sqrt{1-\epsilon} \gg 1$ is fairly satisfied at large distances 
for which the following analytic form of $\xi_{\rm PBH}(r)$ becomes an excellent approximation.

Using the asymptotic formula for ${\rm erfc} (x)$ for 
$x \gg 1$
\begin{equation}
    {\rm erfc} (x) \approx \frac{e^{-x^2}}{\sqrt{\pi}x},
\end{equation}
Eq.~(\ref{P2-2}) becomes
\begin{equation}
     P_2 ({\bm x}_1, {\bm x}_2) \approx
     \frac{1+\epsilon}{2\pi \nu} e^{-\frac{\nu^2}{2}}
     \sqrt{\frac{\pi}{2}} {\rm erfc}
     \left( \sqrt{\frac{1-\epsilon}{2(1+\epsilon)}} \nu \right).
\end{equation}
Since our assumption $\nu \sqrt{1-\epsilon} \gg 1$
automatically means the argument of ${\rm erfc}$ is much
bigger than unity,
we again use the asymptotic formula for ${\rm erfc}$ 
and we arrive at the final expression of $P_2$ as
\begin{equation}
     P_2 ({\bm x}_1, {\bm x}_2) \approx
     \frac{1}{2\pi \nu^2} \frac{{(1+\epsilon)}^{\frac{3}{2}}}{\sqrt{1-\epsilon}}
     \exp \left( -\frac{\nu^2}{1+\epsilon} \right).
\end{equation}
Then, the PBH correlation function becomes
\begin{equation}
\label{xiPBH}
    \xi_{\rm PBH}({\bm x}_1, {\bm x}_2)
    =\frac{P_2({\bm x}_1, {\bm x}_2)}{P_1^2}-1
    \approx 
    \frac{{(1+\epsilon)}^{\frac{3}{2}}}{\sqrt{1-\epsilon}}
     \exp \left( \frac{\epsilon}{1+\epsilon} \nu^2 \right)-1.
\end{equation}
This formula was also obtained in \cite{Ali-Haimoud:2018dau}. Therefore, once we give the two-point correlation function for $\phi$, 
we can obtain the approximate form for the PBH correlation function by this equation.
In the following, we use this equation to compute the PBH correlation function.

\subsection{PBH angular correlation function}
Since the SMBHs at high redshifts (i.e. $z=5 - 7$) distribute inside a (rather) thin
shell in the comoving coordinates where our galaxy is located at the origin,
it will be more convenient to consider the angular correlation function when one
compares the theoretical prediction with the observational data.
Here we summarize the formulas to calculate the angular correlation function of PBHs (for the formalism of the angular correlation function, e.g., see \cite{Peebles:1980}). 

We denote the (comoving) number density of PBHs by $n_{\rm PBH} (R, \theta, \varphi)$ 
where $R, \theta, \varphi$ are the comoving polar coordinate and the observer sits at the origin.  
We also define the two-dimensional number density $N_{\rm PBH} (\theta, \varphi)$ by
\begin{eqnarray}
\label{eq:N_n}
	N_\PBH (\theta, \varphi)=\int_0^\infty W(R)  \,  n_\PBH(R, \theta, \varphi) \, R^2 \, d R \, ,
\end{eqnarray}
where $W(R)$ is the window function which accounts for the redshift range that observations under consideration are sensitive to.
In the following calculation, the window function is chosen such that it corresponds to the distribution of SMBHs observed at high redshifts.
In principle, the window function can also depend on $\theta, \varphi$ as well, but we assume
such dependence is absent in this paper.

The spatial average of $n_{\rm PBH} (R, \theta, \varphi)$ is denoted as $\bar n_{\rm PBH}$, and correspondingly, we can also define the two-dimensional counterpart for $N_{\rm PBH} (\theta, \varphi)$ as 
\begin{eqnarray}
\label{eq:bar_N_n}
	\bar{N}_\PBH=\int_0^\infty   dR \, R^2 \,  W(R) \, \bar{n}_\PBH \,,
\end{eqnarray}
which is the angle averaged number density on the projected two-dimensional sphere.

Now we introduce fluctuations of $n_\PBH (R, \theta, \varphi)$ as 
\begin{eqnarray}
\label{eq:delta_n}
\de_\PBH(R,\theta,\varphi)\equiv\frac{n_\PBH(R,\theta,\varphi)-\bar{n}_\PBH}{\bar{n}_\PBH} \,,
\end{eqnarray}
while fluctuations of the two-dimensional number density $N_{\rm PBH} (\theta, \varphi)$ are defined as 
 \begin{eqnarray}
 \label{eq:delta_2dim_N}
\De_\PBH(\theta,\varphi)\equiv\frac{N_\PBH(\theta,\varphi)-\bar{N}_\PBH}{\bar{N}_\PBH} \,.
 \end{eqnarray}
Since $\bar{N}_\PBH $ is given by Eq.~\eqref{eq:bar_N_n}, 
$\De_\PBH(\theta,\varphi)$ can be written in terms of $\de_\PBH(R,\theta,\varphi) $ as
\begin{eqnarray}
\label{eq:Delta_PBH}
	\De_\PBH(\theta,\varphi)=\int_0^\infty dR \,  g(R) \, W(R) \, \de_\PBH(R,\theta,\varphi),
\end{eqnarray}
where we have defined $g(R)$ as 
\begin{eqnarray}
\label{eq:g_R}
g(R) \equiv \frac{R^2\bar{n}_\PBH}{ \bar N_\PBH} = \frac{R^2}{\displaystyle\int_0^\infty   dR'  R^{\prime 2} \, W(R') } \,.
\end{eqnarray}

The angular correlation function of PBH $w_\PBH (\theta)$ is defined by 
\begin{eqnarray}
 \label{eq:w_PBH_1}
  w_\PBH(\theta )&=&\left\langle  \De_\PBH(\hat r_1 ) \,\, \De_\PBH( \hat r_2) \right\rangle  \,,
\end{eqnarray}
where $\hat r_1 = (\theta_1, \varphi_1)$ and $\hat r_2 =  (\theta_2, \varphi_2)$ correspond to the direction on the sphere and $\theta$ is the angle between $\hat r_1$ and $\hat r_2$. 
By using Eq.~\eqref{eq:Delta_PBH}, one can express $ w_\PBH(\theta )$ in terms of
the PBH correlation function $\xi_\PBH (r)$ as 
 \begin{eqnarray}
\label{eq:w_PBH_2}
  w_\PBH(\theta)
= \int_0^\infty  d  R_1 \int_0^\infty   d R_2  \, g(R_1) \, g(R_2) \, W(R_1) \, W(R_2)  \, \xi_\PBH (r) \,,
 \end{eqnarray}
where $r$ is the covoming distance between positions $\x_1$ and $\x_2$ which is given by 
\begin{eqnarray}
 \label{eq:distance_r}
 r=\sqrt{R_1^2+R_2^2-2R_1R_2 \cos\theta} \,,
\end{eqnarray}
with $R_1 = |\x_1|$ and $R_2 = |\x_2|$ being the comoving distance to positions $\x_1$ and $\x_2$ from an observer at the origin.  The comoving distance $R_i$ to the redshift  $z_i$ can be calculated as
\begin{eqnarray}
R_i (z_i)=\int^{z_i}_{0}\frac{d z}{H(z)} \,,
\end{eqnarray}
where $H(z)$ is the Hubble parameter at the redshift $z$.

In this paper, we consider a flat $\Lambda$CDM model, in which $R_i (z_i)$ can be given by 
\begin{eqnarray}
R_i (z_i)=\int^{z_i}_{0}\frac{d z}{H_0 \sqrt{\Omega_r (1+z)^4 + \Omega_m (1+z)^3 + \Omega_\Lambda}} \,,
\end{eqnarray}
where $\Omega_r, \Omega_m$ and $\Omega_\Lambda$ are density parameters for radiation, matter and a cosmological constant, respectively. 
Unless otherwise stated, we assume $\Omega_m h^2 = 0.1424$ and $h = 0.6766$ \cite{Aghanim:2018eyx} in the following.

With the formalism presented in this section, we discuss the angular correlation function for PBHs in models which can explain SMBHs at high redshifts.

\section{Evaluation of the angular correlation function of PBHs  \label{sec:results}}

Having explained the formalism to compute the PBH angular correlation function $w_{\rm PBH}$ in
the previous section,
in this section we explicitly evaluate the PBH angular correlation function by using
Eq.~\eqref{eq:w_PBH_2} and the PBH correlation function $\xi_\PBH$ given by Eq.~\eqref{xiPBH}.   

First we calculate the PBH correlation function $\xi_\PBH (r)$. 
As we mentioned previously, we consider the scenario where PBHs are formed from fluctuations 
of a light spectator field $\phi$. 
The two-point correlation function of $\phi$ for $r = |\x_1 - \x_2|$ appearing in Eq.~\eqref{xiPBH} should be evaluated at the time of 
the end of inflation $t_{\rm end}$:
\begin{equation}
\label{eq:xi_phi}
\xi_\phi (r) =\left\langle \left(  \phi (t_{\rm end}, \x_1)  - \phi_{\rm ini} \right) \, \left( \phi ( t_{\rm end}, \x_2)  -\phi_{\rm ini} \right) \right\rangle = \int_{k_{\rm min}}^{k_{\rm max}}  \frac{dk}{k} {\cal P}_\phi (k) \frac{\sin (kr)}{kr}  \,.
\end{equation}
Since we take the correlation of $\phi$ inside our observable Universe, the minimum wave number $k_{\rm min}$ is taken to be the present Hubble scale $k_{\rm min} = H_0$ in the following calculations. $k_{\rm max}$  is the maximum wave number corresponding to the mode which gives  the minimum PBH mass to explain SMBHs. In this paper, we assume that SMBHs originate from PBHs with the masses of $M_\PBH \ge 10^4 \, M_\odot$, 
from which $k_{\rm max}$ can be fixed as 
\begin{equation}
\label{eq:k_max}
k_{\rm max} \simeq 1.3 \times 10^4 \, {\rm Mpc}^{-1} \left( \frac{g_\ast}{10.75} \right)^{-1/12} \left( \frac{M_\PBH}{10^4 \, M_\odot} \right)^{-1/2} \,,
\end{equation}
where $g_\ast$ is the effective degrees of freedom, and taken to be $g_\ast = 10.75$ \cite{Kolb:1990vq} 
since PBHs with masses $M_\PBH = 10^4 \, M_\odot$ are formed at around the cosmic time of 1 sec. 

To evaluate $\xi_\phi (r)$, we need the power spectrum of a spectator field $\phi$, 
which is given, for a scalar field with mass $m$ (see. e.g., \cite{Riotto:2002yw}), by
\begin{equation}
\label{phi-P}
 {\cal P}_\phi (k) = \left( \frac{H_I}{2\pi} \right)^2 \left( \frac{k}{a H_I} \right)^{2m^2 /(3H_I^2)}= \left( \frac{H_I}{2\pi} \right)^2 \left( \frac{k}{k_{\rm end}} \right)^{c_I}  \,,
\end{equation}
where $H_I$ is the Hubble scale during inflation (assumed to be constant) and 
$k_{\rm end}$ is the wave number of the mode which exited the horizon at the end of inflation. Taking the reference scale as $k_\ast = 0.002 \, {\rm Mpc}^{-1}$, and the number of $e$-folds from the end of inflation to the time when the mode $k_\ast$ exited the horizon as $N_\ast = 50$, $k_{\rm end}$ is given by $k_{\rm end} \simeq 10^{19} \, {\rm Mpc}^{-1}$.
This value is adopted in the rest of this paper. 
We have also defined the dimensionless parameter $c_I$ for later convenience as 
\begin{equation}
c_I = \frac{2m^2}{3 H_I^2} \,.
\end{equation}
Since we assume a light field for a spectator $\phi$, $c_I$ satisfies $c_I \ll {\cal O}(1)$.

The variance $\sigma^2$ can be calculated from Eqs.~\eqref{eq:xi_phi} and \eqref{phi-P} as 
\begin{eqnarray}
\label{eq:sigma_phi}
\sigma^2 
=  \left( \frac{H_I}{2\pi} \right)^2  \frac{1}{c_I} \left[ \left( \frac{k_{\rm max}}{k_{\rm end}} \right)^{c_I} -  \left( \frac{k_{\rm min}}{k_{\rm end}} \right)^{c_I} \right]  \,.
\end{eqnarray}
Using Eq.~\eqref{eq:xi_phi} and the expression for $\sigma^2$ given by \eqref{eq:sigma_phi}, 
$\epsilon (r)$ defined by Eq.~(\ref{def-epsilon}) can be written as 
\begin{equation}
\label{epsilon-2}
\epsilon (r) = c_I \left[ \left( \frac{k_{\rm max}}{k_{\rm end}} \right)^{c_I} -  \left( \frac{k_{\rm min}}{k_{\rm end}} \right)^{c_I} \right]^{-1} 
 \int_{k_{\rm min}}^{k_{\rm max}}  \frac{dk}{k} \left( \frac{k}{k_{\rm end}} \right)^{c_I}   \frac{\sin (kr)}{kr}  \,.
\end{equation}
Notice that an explicit dependence on the Hubble scale during inflation $H_I$ 
does not appear in $\epsilon (r)$ (hence in $\xi_\PBH (r)$ as well). 
Although the $k$ integration above can be performed numerically, 
it is a numerical obstacle when it is combined with other integrations to compute $w_{\rm PBH}$.   
On the other hand, for $1/(k_{\rm max}r) \ll 1, \,\, k_{\rm min} r \ll 1$,  and  $c_I \ll 1$,
all of which are actually satisfied except for the Hubble scales $r\sim H_0^{-1}$,
by treating them as small quantities,
we can systematically perform the integration analytically order by order. 
To leading order, the result is given by
\begin{align}
\label{eq:2pt_PBH_3}
\epsilon (r) = \left[ \left( \frac{k_{\rm max}}{k_{\rm end}} \right)^{c_I} -  \left( \frac{k_{\rm min}}{k_{\rm end}} \right)^{c_I} \right]^{-1} 
\left[ -  \left( \frac{k_{\rm min}}{k_{\rm end}}\right)^{c_I} 
 + \frac{\cos (c_I \pi / 2) \Gamma (1+ c_I) }{ (1-c_I) (k_{\rm end} r )^{c_I} }
\right] \,, \notag \\
\end{align}
where $\Gamma(x)$ is the Gamma function.
Derivation of this equation and the estimate of its error are given in the appendix~\ref{sec:app}.

\begin{figure}[tbp]
        \begin{center}
          \includegraphics[width=10.0truecm]{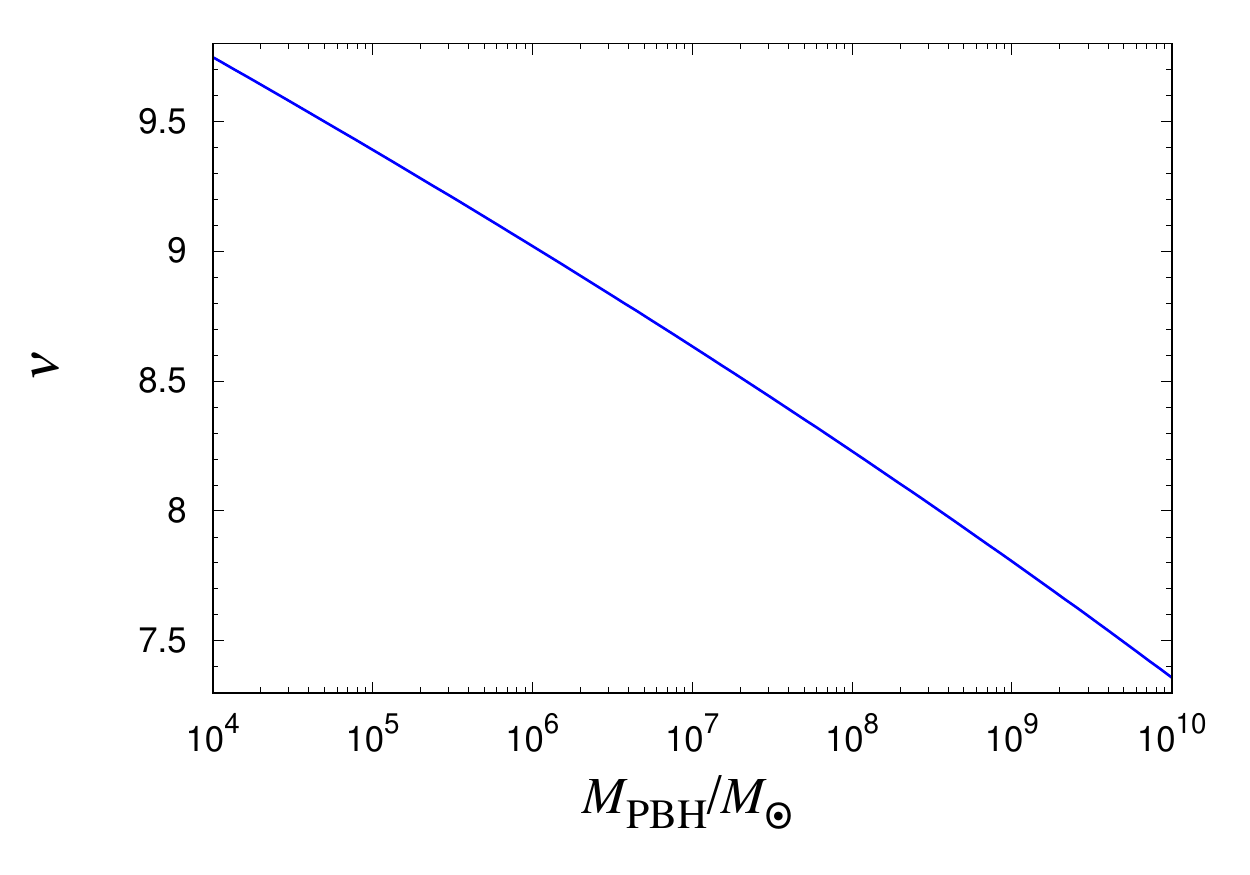}
        \end{center}
      \caption{$\nu$ required to explain the abundance of the observed SMBHs $(n_{\rm PBH} = 1 \, {\rm Gpc}^{-3})$ at high redshifts
      as a function of $M_{\rm PBH}$.}
    \label{fig:nu-mpbh}
\end{figure}

The last ingredient to evaluate the PBH correlation function $\xi_\PBH (r)$
is to fix $\nu$. 
This is done by requiring that the predicted PBH abundance matches
the population of the observed SMBHs. 
The one-point probability of PBHs given in Eq.~\eqref{eq:P_1_3} 
is nothing but the energy fraction of PBHs at their formation time
$\beta$:
\begin{equation}
\label{beta-P1}
\beta = P_1 (\x) = \frac{1}{\sqrt{2\pi} \nu} e^{-\nu^2/2} \,.
\end{equation}
For PBHs with the masses of $M_\PBH$, this can be related to the
comoving PBH number density $\bar{n}_\PBH$ as
\begin{equation}
\label{eq:beta}
\beta \sim 3 \times 10^{-21} \left( \frac{g_\ast}{10.75} \right)^{1/4} \left( \frac{M_\PBH}{10^5 M_\odot} \right)^{3/2} \left( \frac{\bar{n}_\PBH}{{\rm Gpc^{-3}}} \right).
\end{equation}
By equating $\bar{n}_\PBH$ to the comoving number density of the SMBHs observed at high redshifts ( $\sim 1 \, {\rm Gpc}^{-3}$ \cite{Fan:2003wd}),  
we can determine the value of $\nu$ for a given $M_{\rm PBH}$ by the above two equations.
Fig.~\ref{fig:nu-mpbh} shows $\nu$ as a function of $M_{\rm PBH}$.
We confirm that $\nu$ is always much larger than unity
for any values of $M_{\rm PBH}$ in the mass range of our interest.
This is simply because the PBH formation is rare, namely, $\beta \ll 1$.

\begin{figure}[tbp]
        \begin{center}
          \includegraphics[width=10.0truecm]{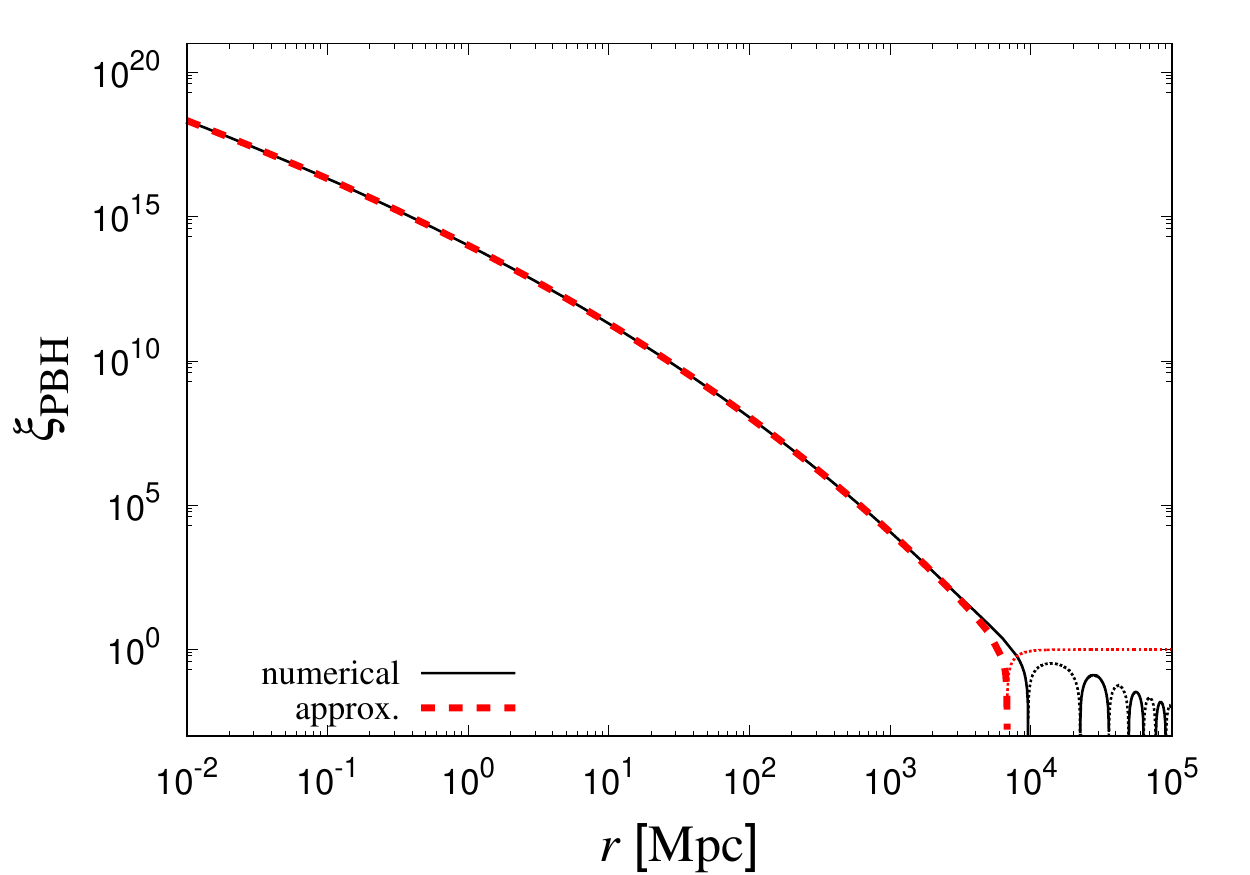}
        \end{center}
      \caption{Comparison of $\xi_\PBH (r)$ between the analytic (approximate) formula based on Eq.~\eqref{eq:2pt_PBH_3} and the numerically obtained one using Eq.~\eqref{epsilon-2}.
      We use Eq.~(\ref{xiPBH}) to translate from $\epsilon (r)$ to $\xi_{\rm PBH}(r)$.
      The values of the model parameters we adopt here are $M_{\rm PBH}=10^4~M_\odot$, $k_\mathrm{end}=10^{19}\ \mathrm{Mpc}^{-1}$, $c_I=0.001$. Dotted curves represent negative values.}
    \label{fig:xi_PBH_analytics_numerics}
\end{figure}

Having clarified how to compute $\epsilon (r)$ and $\nu$,
we can now calculate $\xi_{\rm PBH} (r)$ by using Eq.~(\ref{xiPBH}).
Fig.~\ref{fig:xi_PBH_analytics_numerics} shows two different $\xi_\PBH (r)$: 
the analytic (approximate) one given by Eq.~\eqref{eq:2pt_PBH_3} and the one obtained by the numerical integration of Eq.~\eqref{epsilon-2}. 
As it can be seen from the figure, 
the analytic formula works quite well except for $r$ around the 
Hubble horizon $H_0^{-1}$ where
the neglected components parametrically suppressed by 
${(k_{\rm min}r)}^2$ become important. 
We find that $\xi_{\rm PBH}$ exhibits an
oscillatory feature around $\xi_{\rm PBH}=0$ for large $r$,
which reflects the presence of void regions at such large scales.

Finally, we can calculate the PBH angular correlation function $w_\PBH (\theta)$
by plugging Eq.~\eqref{epsilon-2} [or \eqref{eq:2pt_PBH_3}] into Eq.~\eqref{eq:w_PBH_2}.
To calculate $w_\PBH (\theta)$, we need to determine the form of the window function $W(R)$. 
In the following, we compute $w_\PBH (\theta)$ for two different shapes of the window functions: (i) $\delta$-function form and (ii) top-hat form.  
Since SMBHs are observed in the redshift range of $5 \lesssim z \lesssim 7.5$, the top-hat form would represent a realistic case than the former. 
Yet, to obtain some intuitive understanding by a simple calculation, 
we also consider the  $\delta$-function form.  
For the $\delta$-function case, we assume 
\begin{equation}
\label{eq:W_delta_func}
W  (R(z)) = \delta_{D} (R(z) - R_\ast) \,,
\end{equation}
where we set $ R_\ast= R(z_\ast=7.642)$ where $z_\ast = 7.642$ corresponds to the redshift for the highest quasar  observed at the time of writing this paper \cite{Wang_2021}.  For the top-hat form, the window function is taken as 
\begin{equation}
\label{eq:W_tophat}
W(R(z)) = 
\begin{cases}
1      & (z_{\rm low} \le z \le z_{\rm high}), \\
0    & \text{(otherwise)},
\end{cases}
\end{equation}
where we take $z_{\rm low} = 5$ and $z_{\rm high} = 7.642$ in the following calculations.\footnote{
One may worry that the limit $z_{\rm low} \to z_{\rm high}$ in Eq.~(\ref{eq:W_tophat}) does not reduce to Eq.~(\ref{eq:W_delta_func}).
Such discontinuity is absent in $w_{\rm PBH}$ since $w_{\rm PBH}$
is invariant under the constant multiplication $W(R) \to C W(R)$ 
on the window function [see Eq.~(\ref{eq:w_PBH_2})].
}

Below we show our results for each window function in order.

\subsection{Case of the $\delta$-function window function \label{sec:delta_func}}

First we show the PBH angular correlation function $w_\PBH (\theta)$ 
for the case of the $\delta$-function window function given in Eq.~\eqref{eq:W_delta_func}.  
In this case, $g(R)$ defined in Eq.~\eqref{eq:g_R} becomes unity, i.e. 
$g(R) =1$. Therefore, $w_\PBH (\theta)$ is simply given by 
\begin{equation}
w_\PBH (\theta) = \xi_\PBH (r_\ast) = \xi_\PBH (R_\ast, \theta) \,,
\end{equation}
where $r_\ast$ is given by
\begin{equation}
r_\ast = 2 R_\ast \sin \left( \frac{\theta}{2}  \right)
\end{equation}

\begin{figure}[tbp]
         \begin{minipage}{0.5\hsize}
        \begin{center}
          \includegraphics[width=8.0truecm]{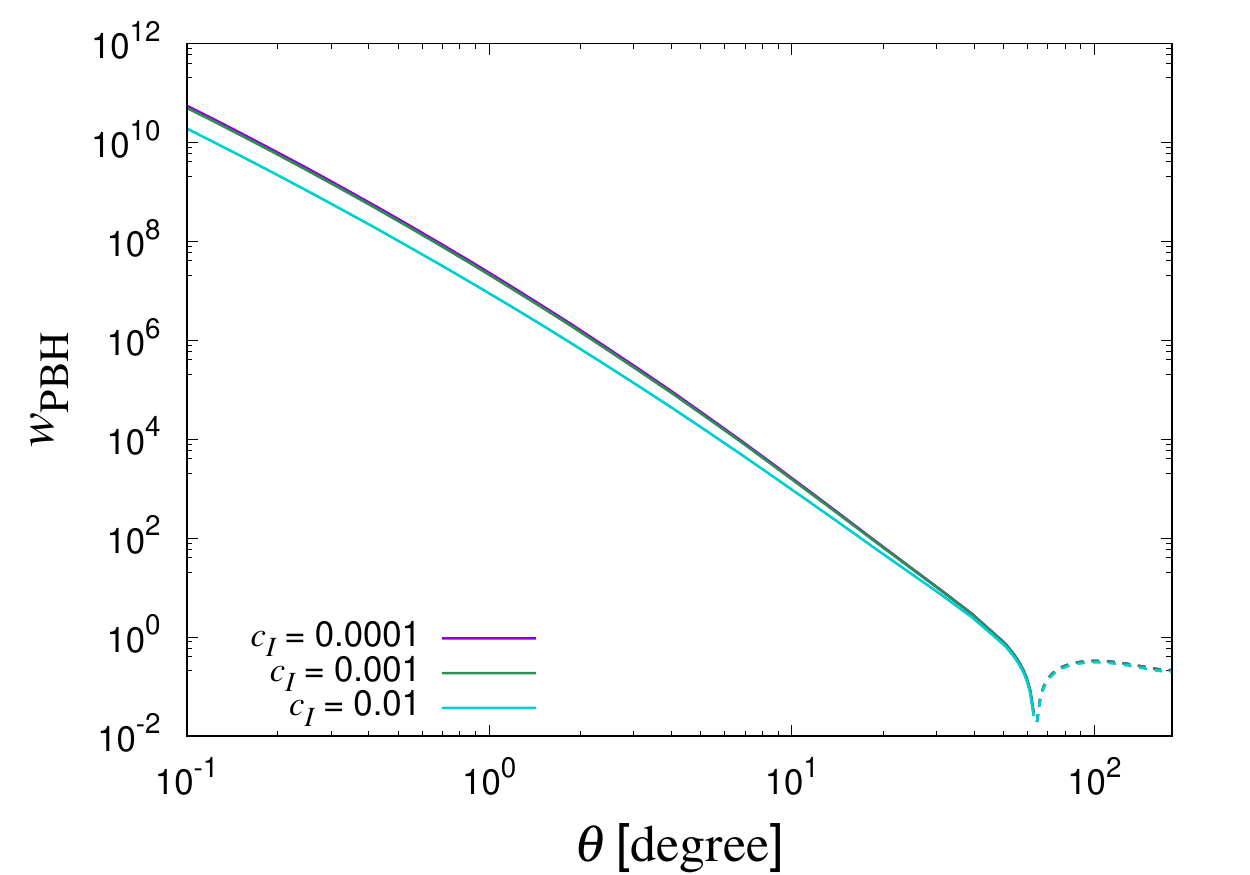}
          \end{center}
          \end{minipage}
         \begin{minipage}{0.5\hsize}
          \begin{center}
          \includegraphics[width=8.0truecm]{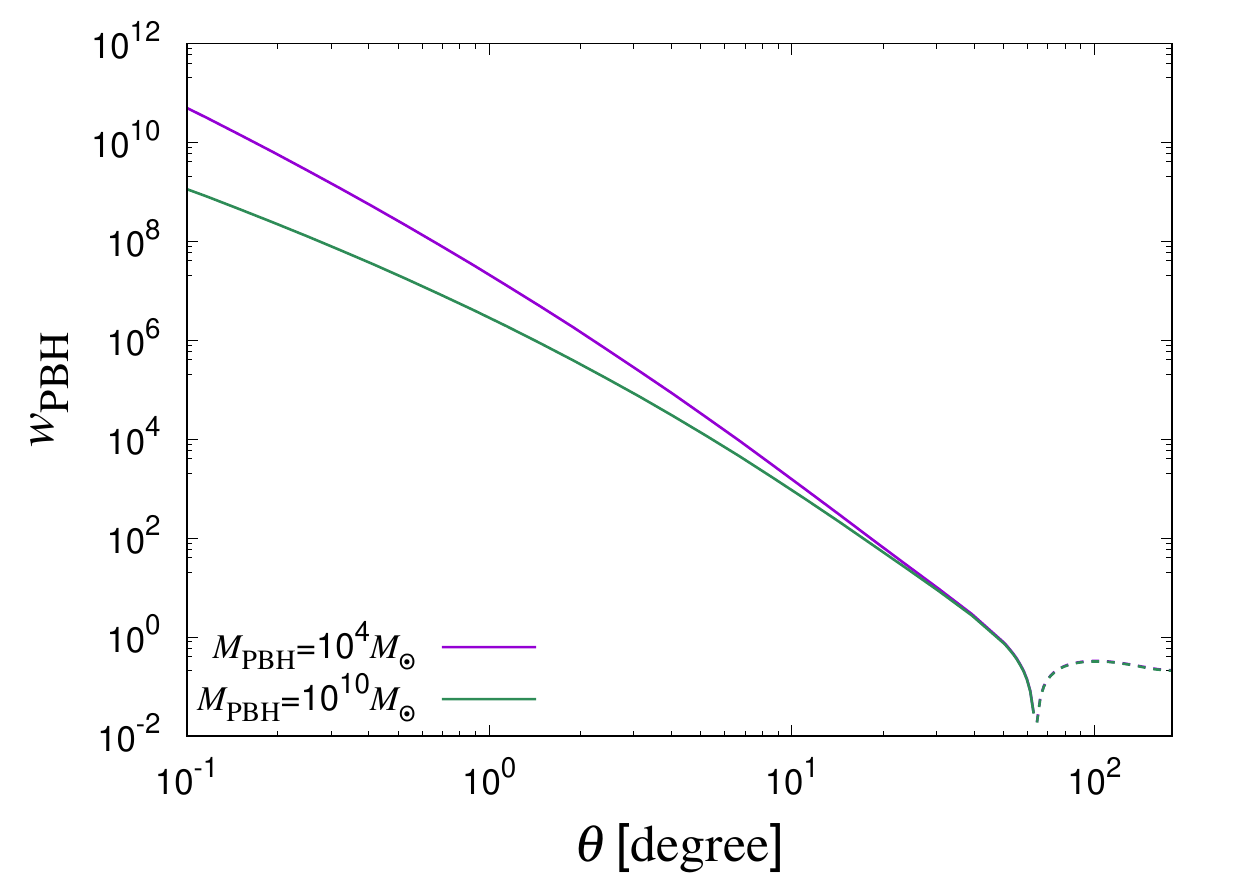}
           \end{center}
          \end{minipage}
      \caption{Angular correlation function $w_\PBH (\theta)$ for the case of the $\delta$-function window at the redshift $z=7.642$. Several cases of $c_I$ are  shown as indicated in the figure. Dotted curves represent negative values.}
    \label{fig:wPBH_delta_func_kmin}
\end{figure}

In Fig.~\ref{fig:wPBH_delta_func_kmin}, we show the PBH angular correlation function for the case with the $\delta$-function window function.  
Here the PBH mass is assumed to be $M_{\rm PBH} = 10^4 M_\odot$. 
For $c_I$, we take it to be $c_I = 0.01, \, 0.001$ and $0.0001$ as shown in the figure.
The right panel shows $w_{\rm PBH} (\theta)$ for $M_{\rm PBH}=10^4 M_\odot$ and $10^{10} M_\odot$.

As seen from the figure, the value of $w_\PBH (\theta)$ is positive and large for 
$\theta < {\cal O}(10^\circ)$, which indicates that PBHs are highly clustered on the sky at those angular scales in models where SMBHs can be explained by PBHs. 
On the other hand, $w_\PBH (\theta)$ for $\theta \gtrsim 60^\circ$ becomes negative
and void regions appear beyond this angle.
This critical angle is neither sensitive to $M_{\rm PBH}$ nor $c_I$.
Thus, the appearance of the void regions for $\theta \gtrsim 60^\circ$ is a robust feature 
which is independent
of the values of the model parameters.
Furthermore, the shape of $w_\PBH (\theta)$ does not much depend on $c_I$, which can also be read off from the figure.  As far as $c_I \ll {\cal O}(1)$, the value of $c_I$ scarcely affects $w_\PBH (\theta)$.  
In summary, PBHs are highly clustered in this kind of PBH model.

\subsection{Case of the top-hat window function \label{sec:top_hat_W}}

When the top-hat window function given as in Eq.~\eqref{eq:W_tophat} is adopted, $g(R)$ can be written by
\begin{eqnarray}
	g(R)=\frac{3R^2}{R_{\rm high}^3-R_{\rm low}^3} \,,
\end{eqnarray}
where $R_{\rm low}$ and $R_{\rm high}$ are the comoving distance to the redshift $z_{\rm low} = 5$ and $z_{\rm high} = 7.642$, respectively. Using this expression for $g(R)$, one can write down $w_\PBH (\theta)$ as
\begin{eqnarray}
\label{eq:wPBH_tophat}
 w_\PBH(\theta )&=& 
 \int_{R_{\rm low}}^{R_{\rm high}}  d  R_1 \int_{R_{\rm low}}^{R_{\rm high}} d R_2  \, \frac{3R_1^2}{R_{\rm high}^3 -R_{\rm low}^3} \, \frac{3R_2^2}{R_{\rm high}^3 -R_{\rm low}^3}
 \, \xi_\PBH (R_1, R_2, \theta) \,,  
\end{eqnarray}
where $\xi_\PBH (R_1, R_2, \theta) \left( = \xi_\PBH (r) \right) $ can be calculated 
by Eq.~\eqref{epsilon-2} [or \eqref{eq:2pt_PBH_3}] with $r$ related to $R_1, R_2$ and $\theta$ by  Eq.~\eqref{eq:distance_r}.

\begin{figure}[tbp]
        \begin{center}
          \includegraphics[width=10.0truecm]{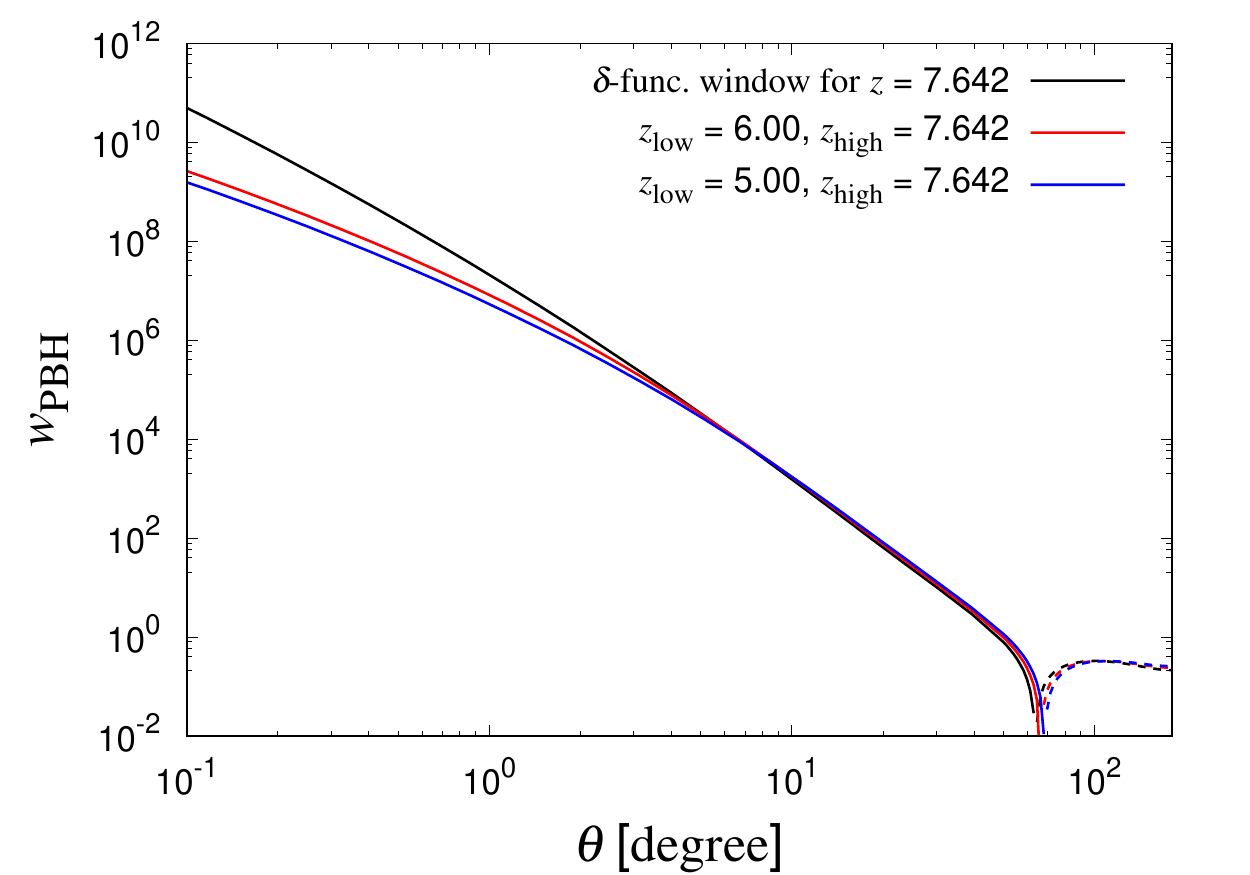}
        \end{center}
      \caption{Angular correlation function $w_\PBH (\theta)$ for the case of the redshfit range of $5 \le z \le 7.642$.  We take $c_I =0.001$ in this figure. Dotted curves represent negative values.}
    \label{fig:wPBH_delta_func_box}
\end{figure}

\begin{figure}[htbp]
        \begin{center}
          \includegraphics[width=10.0truecm]{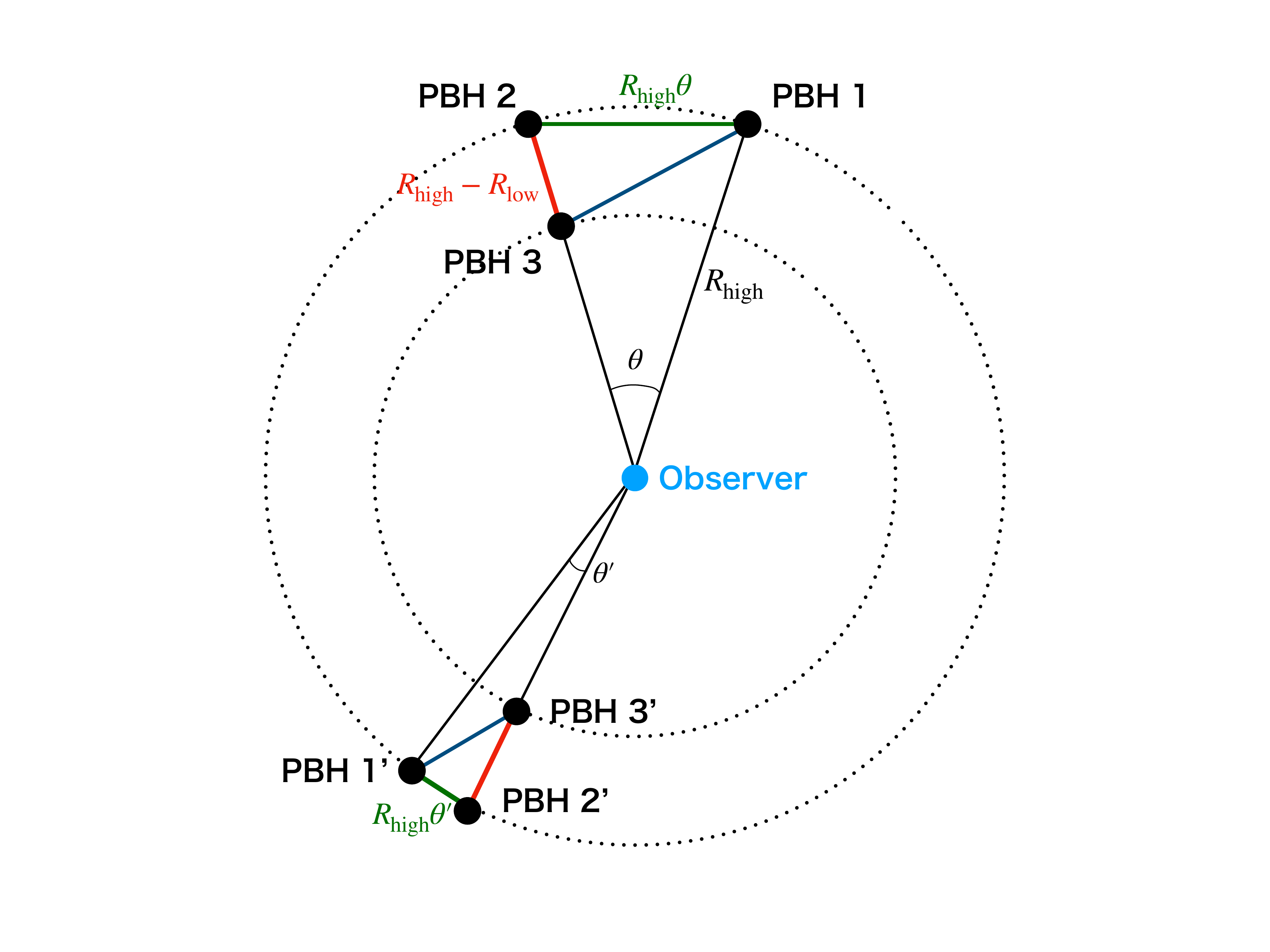}
        \end{center}
      \caption{Schematic explanation of the suppression of $w_\PBH (\theta)$ at small angles
      for the case with the top-hat window function compared to the one for the $\delta$-function window case. }
    \label{fig:2circl}
\end{figure}

\begin{figure}[tbp]
        \begin{minipage}{0.5\hsize}
        \begin{center}
          \includegraphics[width=8.0truecm]{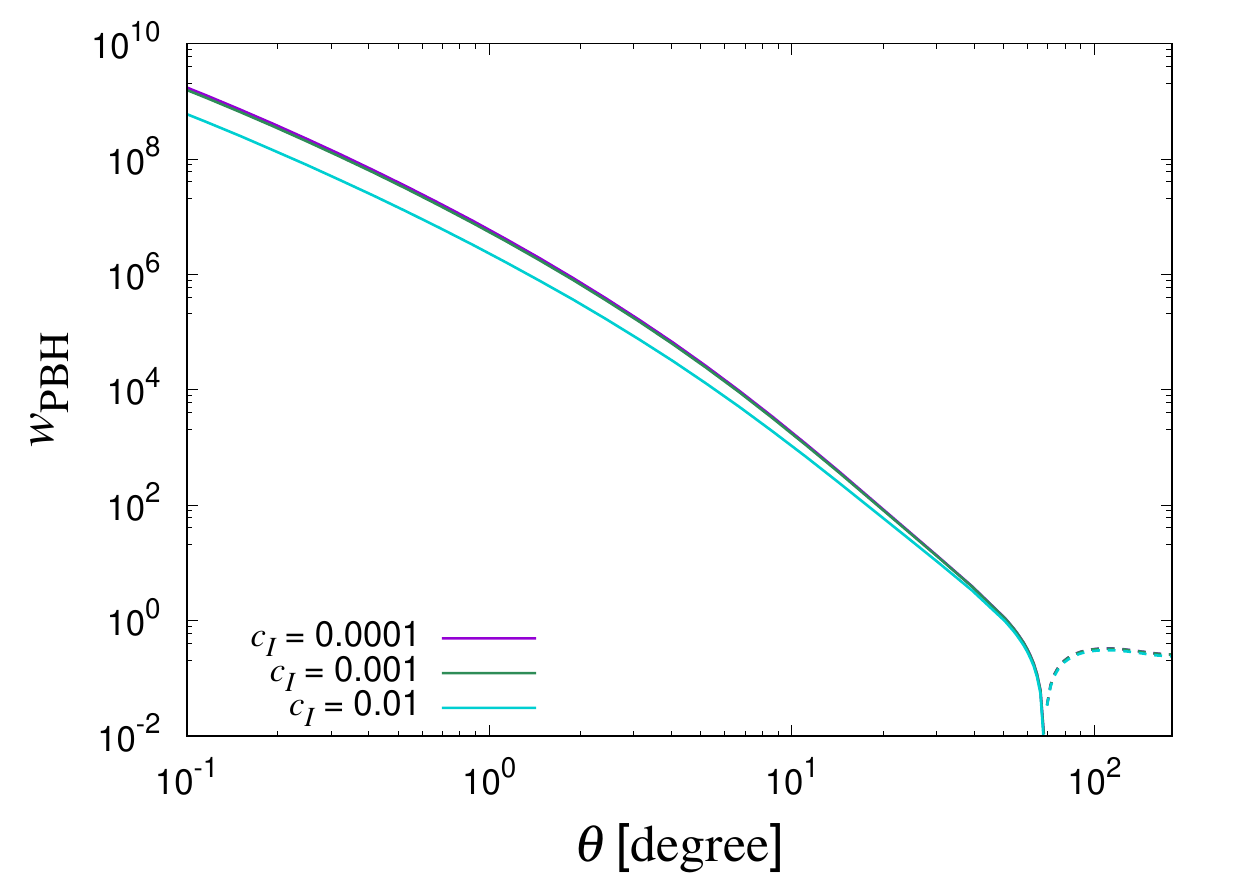}
          \end{center}
          \end{minipage}
         \begin{minipage}{0.5\hsize}
          \begin{center}
          \includegraphics[width=8.0truecm]{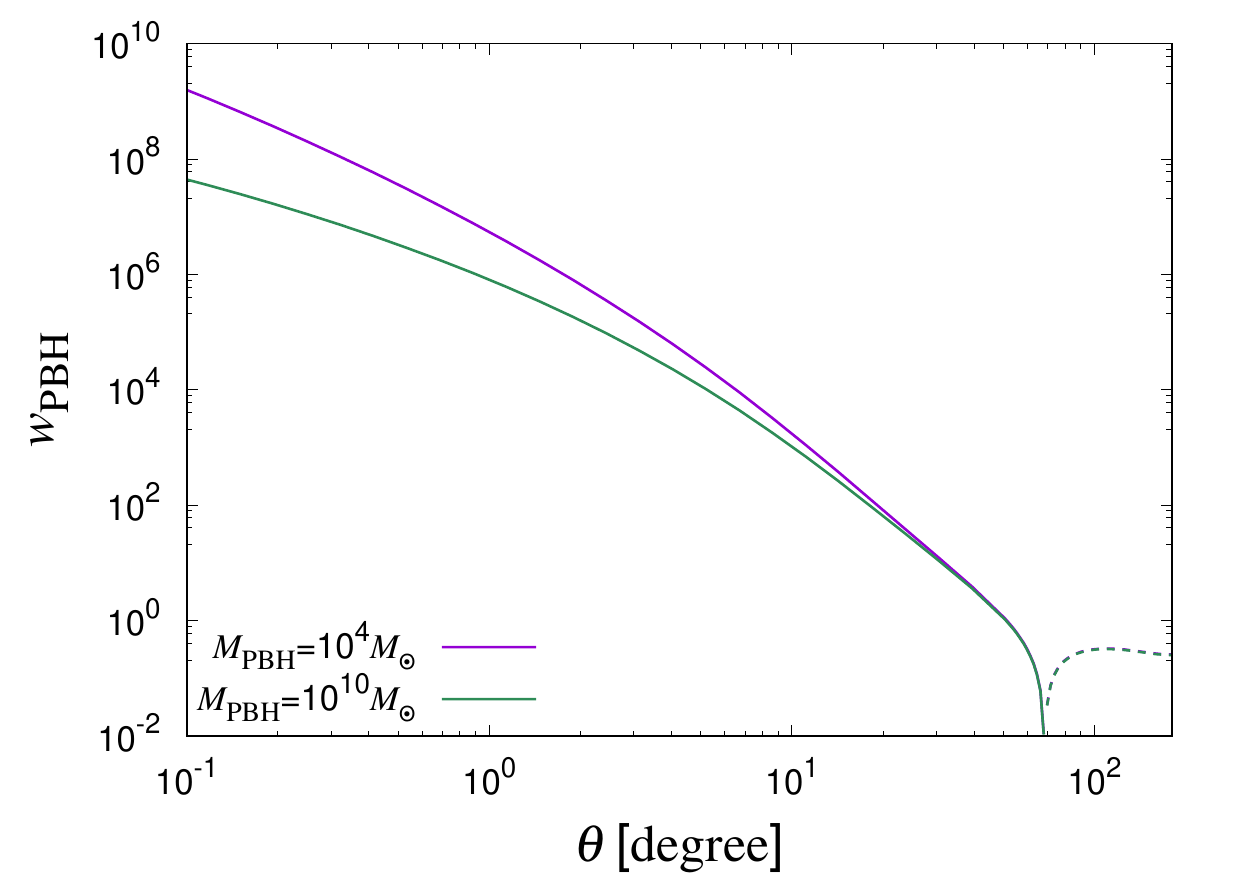}
           \end{center}
          \end{minipage}
              \caption{Angular correlation function $w_\PBH (\theta)$ for the case of the top-hat window function. Several cases of $c_I$ are  shown as indicated in the figure. Dotted curves represent negative values.}
    \label{fig:wPBH_box}
\end{figure}

In Fig.~\ref{fig:wPBH_delta_func_box}, we show the angular correlation function computed from Eq.~\eqref{eq:wPBH_tophat}. For comparison, the case with the $\delta$-function window  is also shown. As seen from the figure, when we assume the top-hat window function,  
$w_\PBH (\theta) $ is reduced on small scales compared to the one for the $\delta$-function window case. 
To understand this feature,
Fig.~\ref{fig:2circl} shows concentric two spheres with radius $R_{\rm low}$ and $R_{\rm high}$ which correspond to the comoving distance to $z_{\rm low}$ and $z_{\rm high}$, respectively, where the observer is located at the origin. 
First of all, it should be noticed that $\xi_\PBH (r)$ is a monotonically decreasing function of $r$ (i.e., the correlation is stronger for shorter distance).  When the interval $R_{\rm high} - R_{\rm low}$ is small  (the limit of $R_{\rm high} - R_{\rm low} \rightarrow 0$ corresponds to the $\delta$-function window case), 
$w_\PBH (\theta)$ measures the correlation of the length scale $R_{\rm high} \theta $ (green line in Fig.~\ref{fig:2circl}). 
On the other hand, if the interval $R_{\rm high} -  R_{\rm low}$  (i.e., red line) is larger than $R_{\rm high} \theta $, the correlation of the length scale $ R_{\rm high} \theta$ contributes less to $w_{\rm PBH} (\theta)$ than the $\delta$-function case.  
Therefore, defining the critical angle $\theta_c$ by
\begin{equation}
\theta_c  \equiv  \frac{R_{\rm high} - R_{\rm low}}{R_{\rm high}} \,,
\end{equation}  
$w_\PBH (\theta )$ for the top-hat window function will be reduced
for $\theta < \theta_c$ compared to the $\delta$-function case. 
For the cosmological parameters adopted in our calculations ($\Omega_m h^2 = 0.1424$ and $h=0.6766$), the comoving distances to $z_{\rm low}$ and $z_{\rm high}$ are given as $R_{\rm low}=7945.9 \, {\rm Mpc}$  and $R_{\rm high} = 9023.1 \, {\rm Mpc}$, respectively, and the critical angle is given by 
\begin{equation}
\theta_c \simeq 7^\circ \,.
\end{equation}  
One can verify in Fig.~\ref{fig:wPBH_delta_func_box} that this critical angle roughly coincides with 
the angle where 
the deviation of $w_\PBH (\theta)$ for the top-hat window case from that for the $\delta$-function case
appears evident. 
We also show dependence on $c_I$ and $M_{\rm PBH}$ of $w_\PBH (\theta)$ for the top-hat window function case in Fig.~\ref{fig:wPBH_box}.  
As in the case for the $\delta$-function window, $w_{\rm PBH}$ is insensitive to $c_I$ but
sensitive to $M_{\rm PBH}$.

\subsection{On the time evolution of the correlation function}
\label{app-fraction}
Since the PBHs in the class of models considered in this paper are highly clustered on small scales,
it is likely that gravitational interactions among neighboring PBHs become non-negligible
and PBHs may undergo frequent mergers or form clumps from the time of the PBH formation
to the time of observations.
However, clarifying the timescale of the PBH mergers requires careful quantitative analysis
since it generally depends on several factors such as the host galaxy shape, rotation and gas content in a complicated manner \cite{Khan:2013wbx, Tremmel:2015rra},
which we do not address in this paper.
Given that $\xi_{\rm PBH} (r)$ and $w_{\rm PBH}(\theta)$ we have computed assume no time evolution
after the PBH formation, direct comparison between such quantities with the observed 
spatial (or angular) distribution of the SMBHs becomes invalid on scales where the correlation function
evolves due to the mergers or the dynamical assembling.
Precise evaluation of this effect requires $N$-body simulations and is beyond the scope of this paper.
Here, by a simple calculation, we estimate the comoving radius 
inside of which the PBHs dominates the mass density.   In such a region,  due to the self-gravitation of PBHs, the clustered PBHs could experience the merger frequently.
At least below this scale, the PBH correlation function at high redshifts ($z=5 - 7$) will significantly differ from the one derived in this paper since we do not take account of the evolution of PBHs such as their merger in the calculation. On the other hand,
we expect that the time evolution of the PBH correlation function for larger scales
is modest although we might need a careful treatment to justify this statement rigorously, which is also beyond the scope of this paper.

Now we estimate the comoving scale under which the mergers of PBHs could occur frequently. Let us suppose that there is one PBH at the origin.
Then, the typical PBH distribution at the comoving distance $r$ from the origin in terms of the mass density is given by
\begin{equation}
    \rho_{\rm PBH} (r)=M_{\rm PBH}(1+\xi_{\rm PBH}(r)) {\bar n}_{\rm PBH}.
\end{equation}
Then, the expected total mass of PBHs inside the radius $r$ becomes
\begin{equation}
    M_{\rm tot}(r)=\int_0^r \rho_{\rm PBH}(r') 4\pi r'^2 dr'
    =\frac{4\pi}{3}M_{\rm PBH} {\bar n}_{\rm PBH}r^3+4\pi M_{\rm PBH}
    {\bar n}_{\rm PBH} \int_0^r r'^2 \xi_{\rm PBH}(r') dr'.
\end{equation}
Thus, the energy fraction of PBHs in total matter inside the radius $r$ becomes
\begin{equation}
\label{fraction}
F(r) \equiv \frac{3M_{\rm tot}(r)}{4\pi r^3 \Omega_{\rm m} \rho_c} \approx 
\frac{3\Omega_{\rm PBH}}{\Omega_{\rm m} r^3}
\int_0^r r'^2 \xi_{\rm PBH}(r')dr' \,,
\end{equation}
where $\rho_c$ is the critical energy density and
\begin{equation}
\label{omega-pbh}
\Omega_{\rm PBH} = \frac{M_{\rm PBH}}{\rho_c V} \simeq 7 \times 10^{-11} \left( \frac{M_{\rm PBH}}{10^{10} M_\odot} \right) \left( \frac{V}{{\rm Gpc}^3} \right)^{-1} \,.
\end{equation}
Here we assume that there is one PBH per ${\rm Gpc}^3$ volume.
When the fraction $F(r)$ is larger than unity, the free-fall time due to self-gravitation of PBHs would be shorter than that of the cosmic expansion, and hence clustered PBHs can attract easily  each other and the mergers may often take place. 

\begin{figure}[t]
  \begin{center}
    \includegraphics[clip,width=10.0cm]{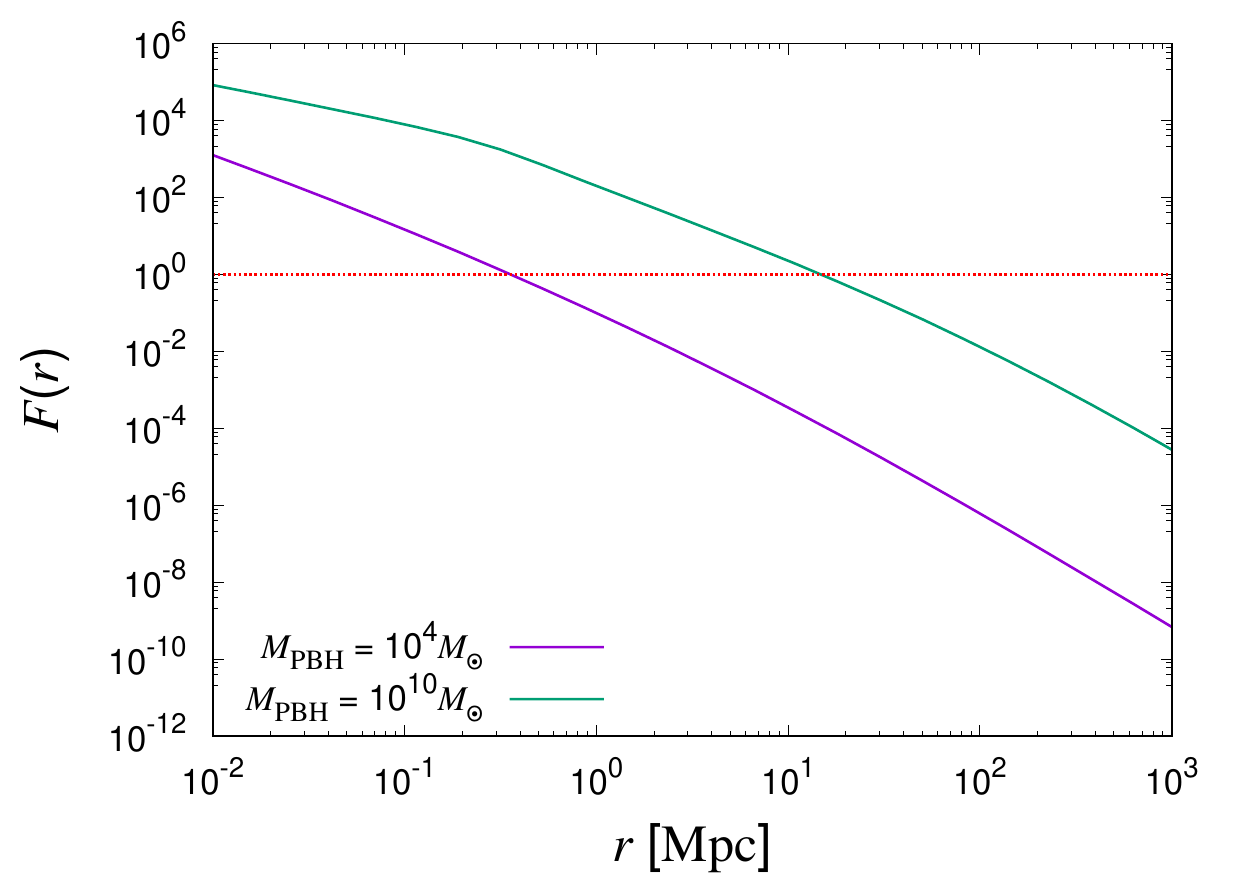}
    \caption{Plot of Eq.~(\ref{fraction}) for two cases: $M_{\rm PBH}=10^4 M_\odot$ and $10^{10}M_\odot$. Red dotted line represents $F(r)=1$. }
    \label{fig-fraction}
  \end{center}
\end{figure}

Fig.~\ref{fig-fraction} shows $F(r)$ for two cases: 
$M_{\rm PBH}=10^4 M_\odot$ and $10^{10}M_\odot$.
We find that $F(r)$ becomes unity at a certain distance in ${\cal O}(0.1) - {\cal O}(10)~{\rm Mpc}$
whose precise value depends on the value of $M_{\rm PBH}$.
The corresponding separation angle evaluated at the redshift $z$ is given by
\begin{equation}
    \theta \equiv \frac{r}{R(z)} \approx 0.07^\circ \left( \frac{r}{10~{\rm Mpc}} \right),
\end{equation}
where we have assumed $z=7$ to obtain the angle on the right-hand side.
Because of this reason, we have set the minimum angle to be $0.1^\circ$ in
all the plots of $w_{\rm PBH}(\theta)$ presented in this section.

As a final remark, we comment that in addition to the large correlation function
of the SMBHs, 
the PBH models we consider may provide another observational signature for gravitational waves (GWs) which
can be a potentially interesting target
for the space interferometers such as LISA \cite{Audley:2017drz}, TianQin \cite{Luo:2015ght} and Taiji \cite{Guo:2018npi} in an optimistic case. 
As we discussed at the beginning of this subsection,
it might be possible that highly clustered PBHs undergo frequent mergers.
If this is true, we can crudely estimate $\Omega_{\rm GW}$ from such PBH mergers
by assuming that the fraction $f$ of PBHs merge as 
\begin{equation}
\label{eq:Omega_GW}
    \Omega_{\rm GW} \simeq 1.5\times 10^{-16} f 
    \left( \frac{\epsilon_{\rm GW}}{0.05} \right)
    {\left( \frac{1+z_{\rm merger}}{10} \right)}^{-1}
        \left( \frac{\Omega_{\rm m}}{0.3} \right)
    \left( \frac{\Omega_{\rm PBH}}{10^{-13}} \right).
\end{equation}
Here $\epsilon_{\rm GW}$ is the fraction of the radiated GW energy 
to the mass of the BH binary \cite{Lousto:2013wta},
and $z_{\rm merger}$ is the redshift at which PBHs merge.
As a fiducial value of $\Omega_{\rm PBH}$, we have adopted $M_{\rm PBH}=10^7~M_\odot$ which corresponds to the maximal mass of the SMBH mergers  detectable by LISA \cite{Audley:2017drz}, 
and used Eq.~(\ref{omega-pbh}) where it is assumed that there is one PBH per ${\rm Gpc}^3$ volume. 
This rough estimation suggests that the GWs from the mergers of the clustered PBHs 
do not exceed the sensitivity reached by LISA \cite{Thrane:2013oya}
even if the large fraction of PBHs [$f={\cal O}(1)$] actually merge.
However, observations of quasars may reveal only a fraction of the SMBHs at high redshifts
and there may be much more lurking SMBHs. In this case, the estimate given in Eq.~\eqref{eq:Omega_GW} may get modified to give a much larger value. 

\section{Conclusion and discussion  \label{sec:conclusion}}

We have investigated the angular correlation function of PBHs as a new probe of the clustering properties of PBHs, 
focusing on a scenario where PBHs can explain SMBHs observed at high redshifts $ 5 \lesssim z \lesssim 7.5$.  
PBHs having initial mass $10^4 M_\odot - 10^{13} M_\odot $ are ruled out from non-observations of the CMB spectral  distortion if they are formed from the Gaussian adiabatic primordial fluctuations. 
However, one can avoid this constraint by considering highly non-Gaussian fluctuations, 
which are subdominant on large scales and scarcely affect observations of CMB anisotopies.

We have considered a class of the spectator field model where the region with high values of $\phi \,\, ( > \phi_c)$ can collapse into PBHs with the masses of $\gtrsim 10^4\, M_\odot$. 
We derived and analyzed the angular correlation function in such a PBH scenario 
and have shown that PBHs are highly clustered,  which suggests that models with PBHs as the origin of SMBHs may be disfavored due to the fact that the clustering is too strong.  This can be confirmed by directly comparing the (theoretically) predicted angular correlation function obtained in this paper with the one from the observed spatial distribution of SMBHs.
In principle, this can be performed by using the existing observations. However, one needs very careful treatment to compare the observed distribution and theoretical prediction obtained in this paper since the quasar searches are not complete, e.g., different observations which found those SMBHs were done under different conditions in terms of the observational time and the area observed.

Once this comparison becomes available, the angular correlation function discussed in this paper should give a critical test for models where PBHs can play a role of SMBH formation, and then one would be able to draw a definite conclusion of whether such a scenario is disfavored or not, which is left for the future project.

\section*{Acknowledgments}
The authors thank Takahiko Matsubara for helpful comments. This work is supported by the MEXT (Ministry of Education, Culture, Sports, Science and Technology) Grant-in-Aid for Scientific Research on Innovative Areas 
No.~17H06359~(T.~Suyama), No.~19K03864~(T.~Suyama), JSPS (Japan Society for the promotion of Science) KAKENHI Grants No. 17H01131~(T.~Takahashi), 19K03874~(T.~Takahashi) and
MEXT KAKENHI Grant No. 19H05110~(T.~Takahashi). 
\\

\appendix

\noindent 
\section{Derivation of Eq.~(\ref{eq:2pt_PBH_3}) \label{sec:app}} 
First of all, since we consider PBHs at high redshift, the comoving distance $R(z)$ can be approximated as 
\begin{equation}
\label{eq:R_approx}
R(z) \simeq R_0 - \frac{2}{H_0/h \sqrt{\Omega_m h^2 ( 1+z)}}  \,,
\end{equation}
where $R_0 \simeq 1.4 \times 10^4 \, {\rm Mpc}$ for $\Omega_m h^2 = 0.1424$ and $h = 0.6766$.  
Therefore, at leading order, one can treat $R(z)$ as constant $R(z) \simeq R_0$ and the 
comoving distance between two points at high $z$ becomes 
\begin{equation}
r \simeq 2 R_0 \sin \left( \frac{\theta}{2} \right) \,,
\end{equation}
where $\theta$ is the angle between the two points measured by the observer located at the origin.
In this case, $k_{\rm max} r $ becomes as 
\begin{equation}
k_{\rm max} r \simeq  3 \times 10^{8} \left( \frac{k_{\rm max}}{10^4 \, {\rm Mpc}^{-1}} \right) \left( \frac{R_0}{1.4 \times 10^4 \, {\rm Mpc}} \right) \sin \left( \frac{\theta}{2} \right) \,.
\end{equation}
The angle scale of our interest from the observational point of view is 
$\theta \ge 0.1^\circ$, for which one can verify that $k_{\rm max}r$ is much larger than unity.
Keeping this in mind, let us decompose the integral of Eq.~\eqref{epsilon-2} as
\begin{equation}
  \int_{k_{\rm min}}^{k_{\rm max}}  \frac{dk}{k} \left( \frac{k}{k_{\rm end}} \right)^{c_I}   \frac{\sin (kr)}{kr}  
 = \int_{k_{\rm min}}^{\infty}  \frac{dk}{k} \left( \frac{k}{k_{\rm end}} \right)^{c_I}   \frac{\sin (kr)}{kr}  
 -\int_{k_{\rm max}}^{\infty}  \frac{dk}{k} \left( \frac{k}{k_{\rm end}} \right)^{c_I}   \frac{\sin (kr)}{kr}.
\end{equation}
The second term on the right-hand side is ${\cal O}( {(k_{\rm max} r)}^{-2} 
{\left( \frac{k_{\rm max}}{k_{\rm end}} \right)}^{c_I} )$ and is negligible for $k_{\rm max} r \gg 1$.
To deal with the first term, we do the integration by parts:
\begin{eqnarray}
\label{eq:k_integral}
 \int_{k_{\rm min}}^{\infty}  \frac{dk}{k} \left( \frac{k}{k_{\rm end}} \right)^{c_I}   \frac{\sin (kr)}{kr}  
 = 
-\frac{1}{c_I} \frac{1}{k_{\rm end}^{c_I}} \left[ 
 k_{\rm min}^{c_I} \frac{\sin (k_{\rm min} r)}{k_{\rm min} r} + \int_{k_{\rm min}}^\infty k^{c_I} \frac{d}{dk} \left(  \frac{\sin (k r)}{k r} \right) dk
\right] \,. \notag \\
\end{eqnarray}
For $c_I \ll 1$ and $k_{\rm min} r \ll 1$, it is convenient to decompose the second term on the
right-hand side as
\begin{equation}
    \int_{k_{\rm min}}^\infty k^{c_I} \frac{d}{dk} \left(  \frac{\sin (k r)}{k r} \right) dk=
    \int_0^\infty k^{c_I} \frac{d}{dk} \left(  \frac{\sin (k r)}{k r} \right) dk-
    \int_0^{k_{\rm min}} k^{c_I} \frac{d}{dk} \left(  \frac{\sin (k r)}{k r} \right) dk. 
\end{equation}
The first term on the right-hand side can be written in terms of the Gamma function as
\begin{equation}
    \int_0^\infty k^{c_I} \frac{d}{dk} \left(  \frac{\sin (k r)}{k r} \right) dk=
    \frac{r^{-c_I} }{-1+c_I} \cos (c_I \pi / 2) \Gamma (1+ c_I).
\end{equation}
As for the second term, expanding the integrand as $\simeq - \frac13 k^{1+c_I}r^2$,
the leading order contribution is given by
\begin{equation}
    \int_0^{k_{\rm min}} k^{c_I} \frac{d}{dk} \left(  \frac{\sin (k r)}{k r} \right) dk=-\frac{k_{\rm min}^{c_I}}{3(2+c_I)}{(k_{\rm min}r)}^2.
\end{equation}
Putting everything together, we finally obtain
\begin{align}
\epsilon (r) = & \left[ \left( \frac{k_{\rm max}}{k_{\rm end}} \right)^{c_I} -  \left( \frac{k_{\rm min}}{k_{\rm end}} \right)^{c_I} \right]^{-1} 
\Bigg[ -  \left( \frac{k_{\rm min}}{k_{\rm end}}\right)^{c_I} 
\left( 1+{\cal O}({(k_{\rm min}r)}^2) \right) \nonumber \\
 &+ \frac{1}{1-c_I} 
 \frac{\cos (c_I \pi / 2) \Gamma (1+ c_I) }{ (k_{\rm end} r )^{c_I} }
 +{\cal O}\left( c_I {(k_{\rm max}r)}^{-2} 
 {\left( \frac{k_{\rm max}}{k_{\rm end}}\right)}^{c_I} \right)
\Bigg].
\end{align}
Dropping the subdominant terms yields Eq.~(\ref{eq:2pt_PBH_3}).

\bibliography{reference_PBH_clustering}

\end{document}